\documentclass[twocolumn, pra, reprint, amsfonts, amsmath, amssymb, superscriptaddress,longbibliography, floatfix]{revtex4-2}
\usepackage{thmtools} 
\usepackage{thm-restate}
\usepackage{amsthm}
\usepackage{lipsum}
\usepackage{amsmath}
\usepackage{graphicx}
\usepackage{amsfonts}
\usepackage{color}
\usepackage{upgreek}
\usepackage{mathtools}
\usepackage{hyperref}
\usepackage{cleveref}
\usepackage{braket}
\usepackage{physics}
\usepackage{tikz}
\usetikzlibrary{shadings, shapes.geometric, calc}
\hypersetup{
    colorlinks = true,
    linkcolor =blue,
	citecolor=blue, 
	urlcolor=blue 
}
\usepackage{array}

\usepackage{braket}
\usepackage{bm}
\usepackage[normalem]{ulem}

\newcommand{\tracedist}[1]{\text{Tracedist}}

\newcommand{\proj}[1]{\ket{#1}\bra{#1}}

\newcommand{\comment}[1]{}
\newcommand{\genrho}[0]{\rho}
\newcommand{\rerho}[0]{\rho}
\newcommand{\cvxrho}[0]{\bar{\rho}}
\newcommand{\vect}[1]{\boldsymbol{#1}}

\theoremstyle{definition}

\newcommand{\mkdev}[1]{\textcolor{black}{#1}}

\begin{document}

\title{Semidefinite programming for understanding the limitations of Lindblad equations}

        \author{Soumyadeep Sarma}
	\email{ssoumyadeep@iisc.ac.in}
        \affiliation{Department of Physics, Indian Institute of Science, Bangalore 560012, India}
	\affiliation{International Centre for Theoretical Sciences, Tata Institute of Fundamental Research, Bangalore 560089, India}

\author{Manas Kulkarni}
	\email{manas.kulkarni@icts.res.in}
	\affiliation{International Centre for Theoretical Sciences, Tata Institute of Fundamental Research, Bangalore 560089, India}

    \author{Archak Purkayastha}
\email{archak.p@phy.iith.ac.in}
\affiliation{Department of Physics, Indian Institute of Technology, Hyderabad 502284, India}

	\author{Devashish Tupkary}
	\email{djtupkary@uwaterloo.ca}
	\affiliation{Institute for Quantum Computing and Department of Physics and Astronomy, University of Waterloo, Waterloo, Ontario, Canada, N2L 3G1}

\date\today

\begin{abstract}
       

Lindbladian quantum master equations (LEs) are the most popular descriptions for quantum systems weakly coupled to baths. But, recent works have established that in many situations such Markovian descriptions are fundamentally limited: they cannot simultaneously capture populations and coherences even to the leading-order in system-bath couplings. This can cause violation of fundamental properties like thermalization and continuity equations associated with local conservation laws, even when such properties are expected in the actual setting. This begs the question: given a physical situation, how do we know if there exists an LE that describes it to a desired accuracy? Here we show that, for both equilibrium and non-equilibrium steady states (NESS), this question can be succinctly formulated as a semidefinite program (SDP), a convex optimization technique. If a solution to the SDP can be found to a desired accuracy, then an LE description is possible for the chosen setting. If not, no LE description is fundamentally attainable, showing that a consistent Markovian treatment is impossible even at weak system-bath coupling for that particular setting. Considering few qubit isotropic XXZ-type models coupled to multiple baths, we find that in most parameter regimes, LE description giving accurate populations and coherences to leading-order is unattainable, leading to rigorous no-go results. However, in some cases, LE description having correct populations but inaccurate coherences, and satisfying local conservation laws, is possible over some of the parameter regimes. Our work highlights the power of semidefinite programming in the analysis of physically consistent LEs, thereby, in understanding the limits of Markovian descriptions at weak system-bath couplings.
\end{abstract}

\maketitle
	

\section{Introduction}
{\it Lindblad quantum master equations ---}With the advent of quantum technology, accurately describing small quantum systems out of equilibrium has become paramount. However, this is a challenging theoretical problem in general, even when the number of degrees of freedom in the system is small. Consequently, several approximations are very often required \cite{breuer_book, carmichael_book, RivasHuelga2012, Campaioli_2024, Stefanini_2025}. One of the most common approximations, which greatly simplifies such descriptions, is to assume that the dissipative dynamics of the system is Markovian. Indeed, Markovian quantum master equations form the backbone of much of modern open quantum system dynamics, capturing the evolution of systems coupled to external baths in both equilibrium and non-equilibrium conditions \cite{breuer_book, carmichael_book, RivasHuelga2012, Campaioli_2024, Stefanini_2025}, as well as shaping our understanding of how quantum measurements are performed in practice \cite{wiseman_milburn_book, Milburn_book1, Nielsen_and_Chuang, Hacohen-Gourgy_2020, Gambetta_2008, Carmichael_1993}. It was shown by Gorini, Kossakowski, Sudarshan, and Lindblad (GKSL) \cite{GKS1976, Lindblad1976} that any quantum master equation that preserves complete positivity and trace of the density matrix, and describes Markovian dynamics with no explicit time-dependence has to be of the form 
\begin{equation}
\begin{aligned}
	\label{Lindblad_form}
	& \frac{\partial \rho}{\partial t}=i[\rho,H_S + H_{LS}]+\mathcal{D}(\rho), \\
	&\mathcal{D}(\rho) = \sum_{\lambda=1}^{d^2-1}\gamma_{\lambda} \Big( L_{\lambda} \rho L_{\lambda}^\dagger - \frac{1}{2} \{ L_{\lambda}^\dagger L_{\lambda}, \rho   \} \Big),~~\gamma_\lambda \geq 0,
\end{aligned}
\end{equation}
which is commonly called the Lindblad quantum master equation (LE). In Eq.~\eqref{Lindblad_form},  $\rho$ is the density matrix of the system, $d$ is the Hilbert space dimension of the system, $H_S$ is the system Hamiltonian, $H_{LS}$ is the so-called Lamb shift Hamiltonian, $L_\lambda$ are the Lindblad operators, $\gamma_\lambda$ are the rates, and $\mathcal{D}(\rho)$ is called the ``dissipator" term. The preservation of complete positivity condition is enforced by demanding $\gamma_\lambda \geq 0$. Within the Markovian regime, the GKSL framework~\cite{Lindblad1976, GKS1976} guarantees complete positivity and trace preservation (CPTP), ensuring physical consistency.

However, Eq.\eqref{Lindblad_form} only gives the mathematical form of the Markovian quantum master equation. The Hamiltonians, the Lindblad operators and the rates need to be found based on the physical situation to be described. In practice, the system Hamiltonian may be known, but at best partial information is available for the baths, like their spectral functions, and thermodynamic properties like temperatures and chemical potentials. Given such physical situations, it becomes important to find conditions under which Eq. \eqref{Lindblad_form} may be obtained.

{\it Born-Markov approximation and Redfield equation ---}
The most standard and general way to arrive at Markovian quantum master equations in such microscopic approach is via the Born-Markov approximation \cite{breuer_book, carmichael_book, RivasHuelga2012, Campaioli_2024, Stefanini_2025}. This relies on weak system-bath couplings and a time-scale separation between system and the baths, obtaining a quantum master equation to leading-order in system-bath coupling. The quantum master equation obtained is of the so-called Bloch-Redfield form \cite{Bloch_1957, Redfield_1957}, commonly called Redfield equation (RE) 
\mkdev{\begin{align}
	\label{TCL2_intro}
	\frac{\partial{\rerho}}{\partial t}&= i [\rerho(t), H_S]+ \epsilon^2 \mathcal{L}_{2}[\rerho(t)],  \\
    \mathcal{L}_2[\rerho(t)]&= \int_0^\infty dt^\prime \operatorname{Tr}_B [H_{SB},[H_{SB}(-t^\prime),\rerho(t) \otimes \rerho_B]] \notag,
\end{align}
with $H_{SB}(t)$ is the system-bath coupling operator time evolved in the interaction picture, and $\rho_B$ being the initial state of the bath.} After some algebra, the RE can be cast into the form of the LE [Eq. \eqref{Lindblad_form}], with one major caveat. It can be shown in generality that, unless in very special situations, one or more of the rates, $\gamma_{\lambda}$, will be negative \cite{Suarez_1992, Gnutzmann_1996, Hartmann_2020_1, tupkary_searching_2023}. This violates the complete positivity requirement. So, time evolution by RE may cause the density matrix $\rho$ to have unphysical negative eigenvalues. To alleviate this issue, various further approximations are often performed to make the rates positive, leading to various forms of LE  \cite{Rivas_2010,Trushechkin_2016,Decordi_2017,Kirvsanskas_2018, Cattaneo_2019,Davidovic_2020,Mozgunov_2020,Mccauley_2020,Kleinherbers_2020, Nathan_2020,Potts_2021, Scali_2021, Becker_2021,Davidovic_2022,Trushechkin_2022,Nathan_2024, Schnell_2025,Pyurbeeva_2026}.

{\it The LE versus RE conundrum ---}
Despite not having complete positivity, remarkably, the RE has been rigorously shown to give both populations (diagonal elements of $\rho$ in eigenbasis of $H_S$) and coherences (off-diagonal elements of $\rho$ in eigenbasis of $H_S$) correctly to leading-order in system-bath couplings \cite{Tupkary_2022, fleming_cummings_accuracy}. As a consequence, the RE has been shown to perfectly satisfy local conservation laws of the actual physical situation and capture thermalization when that is physically expected \cite{Tupkary_2022}. Contrarily, it has been recently argued rigorously that most further approximations to re-establish complete positivity will lead to inaccurate populations or coherences even in the leading-order. \mkdev {This may cause violation of continuity equations related to absence of local conservation laws and of thermalization, even when such properties are expected in the actual physical situation \cite{Tupkary_2022}.}

This leads to a conundrum. It shows that even at weak system-bath coupling, a consistent Markovian quantum master equation description, which can accurately give the density matrix to the leading-order in system-bath coupling, is hard to find in most situations. One is forced to either choose RE, giving up complete positivity, or choose one of the forms of LE, allowing violation of some fundamental properties and inaccuracies in the density matrix even to the leading-order. 

{\it Present paper: semidefinite program to check for possible LE---}
The above general discussion leads to the following question: given a physical situation, how do we know if an LE description with desired accuracy is attainable that does not violate any fundamental property of the setting? The fact that there are various different approximation schemes to recover an LE from RE \cite{Rivas_2010,Trushechkin_2016,Decordi_2017,Kirvsanskas_2018, Cattaneo_2019,Davidovic_2020,Mozgunov_2020,Mccauley_2020,Nathan_2020,Davidovic_2022, Trushechkin_2022, Scali_2021, Becker_2021, Schnell_2025} makes answering this question difficult, especially when the answer is negative. For a given physical setting, even if several approximation schemes have been tried and none satisfy the desired requirements, one cannot easily rule out the possibility of an improved approximation scheme which meets all requirements for a consistent Markovian description. 

In the present paper, we seek to address this. For a non-degenerate finite-dimensional system Hamiltonians, and in absence of any explicit time-dependent drive, we provide a numerical technique which can rigorously show whether a consistent LE description of a non-equilibrium steady state (NESS) is possible for a given physical situation, up to a chosen accuracy. We do so by showing that the above question can be formulated as a convex optimization problem in the form of a semidefinite program (SDP)~\cite{SDP_review}. The SDP can be solved using standard packages in high-level programming languages. Crucially, when the resulting optimization problem takes the form of a semidefinite program (SDP), its well-understood duality structure allows one to obtain rigorous and tight bounds on the optimal value, thereby providing definitive statements about whether the desired properties can be achieved up to a prescribed tolerance. If a solution can be found to a desired accuracy, then an LE description is possible. Standard solvers also output one candidate for such an LE. If the SDP does not have a solution to our desired accuracy, an LE description is fundamentally unattainable, thereby giving a rigorous no-go result for any Markovian description of that physical setting. Such a fundamental lack of Markovian description means the dynamics necessarily have some non-Markovian features.

Even if there is no LE possible that correctly gives both populations and coherences to leading-order, an LE giving only correct populations and incorrect coherences (thereby having leading-order inaccuracies in density matrix) may still be possible. We show that such LEs can also be searched for by casting the problem into a slightly different SDP.

We remark here that while SDPs are widely used in quantum information and communication, only few previous works have used it for open quantum system descriptions \cite{Wolf_2008, Cubitt_2012, Kiilerich_2018, tupkary_searching_2023}. In Ref.~\onlinecite{Wolf_2008}, the problem of assessing non-Markovianity of a quantum channel was converted into an SDP, while in Ref.~\onlinecite{tupkary_searching_2023}, some of the present authors formulated the problem of searching for LEs showing thermalization and satisfying local conservation laws into an SDP. Our present results stem from combining the ideas in Ref.~\onlinecite{tupkary_searching_2023} with the fact the RE gives NESS populations and coherences in energy eigenbasis correctly to the leading-order \cite{Tupkary_2022, fleming_cummings_accuracy}. One major difference is that in this work we study both equilibrium and non-equilibrium settings, and investigate a broader class of scenarios. This  builds on and utilizes the results of Ref.~\onlinecite{tupkary_searching_2023}, which focused exclusively on the equilibrium case.

We exemplify our methods by applying to few-qubit isotropic XXZ-type chains coupled to multiple baths. Below we summarize our results in this regard.
\begin{enumerate}
	\item {\it First and last qubits attached to baths at finite temperatures:} In this case, LE description is fundamentally unattainable across all system parameter regimes. This is true at equilibrium (both baths having same temperature), as well as at NESS (both baths having different temperatures). It is even impossible to have an LE description having only correct populations (incorrect coherences) to leading-order and satisfying local conservation laws.
	
	\item {\it First two and last two qubits attached to baths at finite temperatures:} In this case also, LE description is fundamentally unattainable across all parameter regimes. However, when all qubits are identical, and the coupling between qubits is relatively low, an LE description giving only correct steady-state populations (but incorrect coherences) and satisfying local conservation laws is possible. This is also true both in equilibrium (i.e, when all baths have same temperature) and in NESS.
	
	\item {\it Finite temperature baths attached at all the qubits:} In the equilibrium case in this setting, a physically consistent LE {\it is attainable}. However, at NESS, when temperature of even one bath is different, or even one site is detached from a bath, LE description is unattainable. 
\end{enumerate}
The above gives rigorous no-go results for Markovian description of steady-state of dissipative isotropic XXZ-type qubit chains, even when the baths are weakly coupled.

The remainder of the paper is structured as follows: In Sec.~\ref{sec:sec2}, we explain the system and heat bath models\mkdev{, alongside the assumptions required to solve for the steady-state in an order-by-order in fashion, which is then explained in Sec.~\ref{sec1-1}. We then review the microscopically derived Bloch-Redfield equation, which is a truncated form of such an expansion in Sec.~\ref{sec1-2}. Sec.~\ref{sec1-3} focuses on the general requirements in Markovian QME and then in Sec.~\ref{sec1-4}, we define the metrics $\tau^{\rm pop}$ (for accurate leading-order populations) and $\tau^{\rm pop,coh}$ (for accurate leading-order populations and coherences) and show how these can be optimized as semidefinite programs}. We also highlight our analytically calculated lower bound on the trace distance between the zeroth-order NESS and the zeroth-order NESS obtained from any LE. In Sec.~\ref{sec:sec3}, we present the main numerical results of our optimized metrics plotted over a large parameter regime summarised in the form of Table~\ref{Table:figs_main} and discuss the differences between cases of attaching one and two qubits to the left and right baths. We also present numerical data and discuss how attaching all qubits to baths in equilibrium leads to a NESS with correct leading-order populations and coherences. We conclude with a summary and outlook in Sec.~\ref{sec:discussion} and delegate important technical details to the appendices.

\mkdev{\section{Model description and assumptions}
\label{sec:sec2}
In this section, we start with a description of our system and then discuss the assumptions we use throughout this work to solve for the steady-state.}

\subsection{System and heat bath description}
\begin{figure}
	\centering
	\includegraphics[width=\columnwidth]{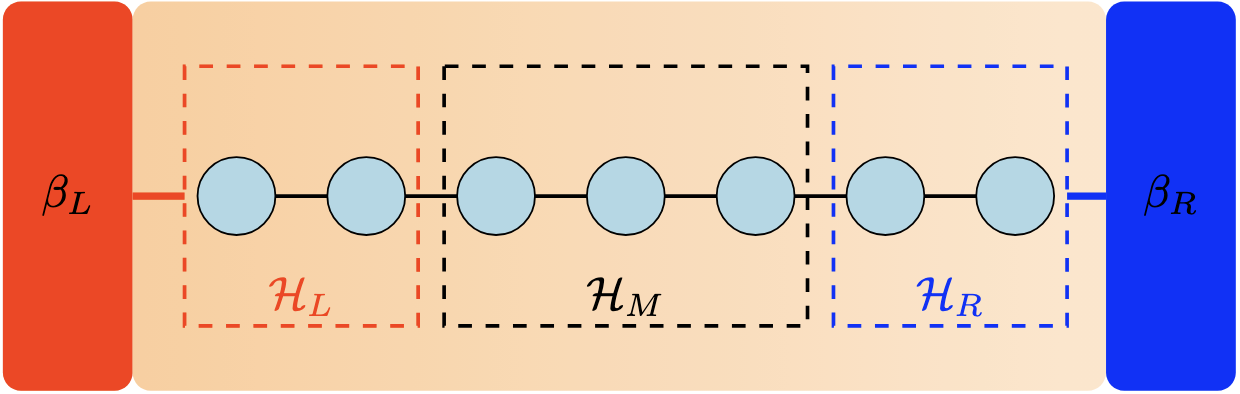}
	\caption{Schematic of an arbitrary finite-dimensional system described by Hamiltonian $ H_S$ [Eq.~\eqref{eq:system_ham}], parts of which are weakly coupled to left and right thermal baths [Eq.~\eqref{eq:bath_ham}] at inverse temperatures $\beta_L$ and $\beta_R$. The Hilbert space of the system, $\mathcal{H}_S$, is divided into parts $\mathcal{H}_L$ (which directly couples to the Hilbert space of the left bath), $\mathcal{H}_R$ (which directly couples to the Hilbert space of the right bath), and the remaining part $\mathcal{H}_M$.}
	\label{fig:main-img}
\end{figure}
Here, we consider the following generic system-bath setup \mkdev{as shown in Fig.~\ref{fig:main-img}}. Let us denote $\mathcal H_L~(\mathcal H_R)$ as the Hilbert space of the part of the system that connects to the left (right) bath and let $\mathcal H_M$ be the Hilbert space of the remaining part of the system. The Hamiltonian ${H}$ of the full setup including the system and all the baths is [see Fig.~\ref{fig:main-img}]
\begin{align}  \label{tot_ham}
	{H}&={H}_S + \epsilon {H}_{SB} + {H}_B, \\ \label{eq:system_ham}
	{H}_S &=  H_L +  H_{LM} +  H_M +  H_{RM} +  H_R, \\ \label{eq:bath_ham}
	H_B &=  H_B^L +  H_B^R.
\end{align}
Here ${H}_B$ is the composite Hamiltonian of all the baths, composed of Hamiltonian for the left ($ H_B^L$) and right ($ H_B^R$) baths with inverse temperatures $\beta_L$ and $\beta_R$ respectively. ${H}_{SB}$ is the composite Hamiltonian describing system-bath coupling to all the baths and $\epsilon\ll 1$ is a dimensionless parameter that controls the strength of system-bath coupling. The Hamiltonian $ H_L~( H_R)$ acts on the Hilbert space $\mathcal H_L~(\mathcal H_R)$, the Hamiltonian $ H_M$ on the Hilbert space $\mathcal H_M$, and $H_{LM}~(H_{RM})$ on both Hilbert spaces $\mathcal H_L$ ($\mathcal{H}_R$) and $\mathcal H_M$.

The initial state of the full set-up is taken as $\genrho_{\rm tot}(0)=\genrho(0) \otimes \genrho_B$, where $\genrho(0)$ is the initial state of the system and $\genrho_B$ is the composite state of all the baths. Without loss of generality, it is possible to assume that ${\rm Tr}_{\rm B}\left({H}_{SB}~ \genrho_B\right)=0$~\cite{breuer_book}. The state of the system at a time $t$ is 
\begin{align}\label{tracefin}
	\genrho(t) = \text{Tr}_{\rm B}\left(e^{-i{H}t} \genrho(0)\otimes\genrho_B e^{i{H}t}\right),
\end{align}
where $\text{Tr}_{\rm B}(...)$ implies trace over bath degrees of freedom. Notice that Eq.~\eqref{tracefin}, by construction, is a CPTP map from $\genrho(0)$ to $\genrho(t)$ \cite{breuer_book,Rivas_2012}. We will assume that in the long-time limit, the system reaches a unique NESS. This assumption physically necessitates that the system size is finite while the baths are in the thermodynamic limit. For simplicity, we will further assume that there are no degeneracies in ${H}_S$. This prevents coherences from possibly having $\mathcal O(1)$ contributions. We shall see later that leading-order contributions for coherences is $\mathcal O(\epsilon^2)$. The NESS density matrix is then defined as
\begin{align}
	\genrho_{\rm NESS} = \lim_{t\rightarrow \infty} \text{Tr}_{\rm B}\left(e^{-i{H}t} \genrho(0)\otimes\genrho_B e^{i{H}t}\right).
\end{align}
It is possible to expand the NESS density matrix in powers of $\epsilon$. Owing to a \mkdev{generalisation of the previously mentioned assumption that ${\rm Tr}_{\rm B}\left({H}_{SB}~ \genrho_B\right)=0$, to a partial trace over bath variables. The term inside the trace is a product of system-bath Hamiltonians of an odd number of terms
\begin{align} \label{eq:odd_point}
 \text{Tr}_{\rm B}\left(H_{SB}(t_1)H_{SB}(t_2)\cdots H_{SB}(t_{2n + 1})\rho_B\right)= 0,
\end{align}}
\mkdev{Hence, with the given Hamiltonian description as in Eqs.~\eqref{tot_ham} to \eqref{eq:bath_ham}, and the generalised assumption in Eq.~\eqref{eq:odd_point}, we proceed to deriving the steady-state in orders of system-bath coupling.}
\mkdev{\section{Solving for steady-state order-by-order}
\label{sec1-1}
In this section, we highlight an expansion of the exact steady-state in powers of the system-bath coupling $\epsilon$. 
Assuming Eq.~\eqref{eq:odd_point} holds,} it can be shown that only even powers of $\epsilon$ can feature in any such expansion, \mkdev{utilizing the time-convolution-less form (TCL) [Secs.~3.3,~9.2.2 of Ref.~\onlinecite{breuer_book}]}. Hence we have,
\begin{align}
	\label{NESS_expansion}
	\genrho_{\rm NESS} = \sum_{m=0}^{\infty} \epsilon^{2m} \genrho_{\rm NESS}^{(2m)},
\end{align} 
where $\genrho_{\rm NESS}^{(2m)}$ stands for the order-by-order NESS density matrix.
The QME describing the evolution of $\genrho(t)$ in the long-time limit can also be expanded in the TCL form. Such an expansion up to some arbitrary high-order, can be formally written as
\begin{align}
	\label{general_TCL_QME}
	\frac{\partial\genrho}{\partial t}= \sum_{m=0}^{\infty} \epsilon^{2m} {\mathcal{L}}_{2m}[\genrho(t)].
\end{align} 
In Eq.~\eqref{general_TCL_QME}, {for $m=0$, }we have that:
\begin{align}
	\label{L0}
	{\mathcal{L}}_{0}[\genrho(t)]=i[\genrho(t),{H}_S]. 
\end{align}
The higher order superoperators $ {\mathcal{L}}_{2m}[\genrho(t)]$ that appear in Eq.~\eqref{general_TCL_QME} are more complicated and there is a systematic way to obtain them. Because the QME must be a linear equation, each superoperator appearing in the above expansion is linear.
Since, $\genrho_{\rm NESS}$ is the long-time steady-state, by definition, it satisfies
\begin{align}
	\label{exact_NESS_equation}
	0=\sum_{m=0}^{\infty} \epsilon^{2m} {\mathcal{L}}_{2m}[\genrho_{\rm NESS}].
\end{align}
Using the expression for $\genrho_{\rm NESS}$ from Eq.~\eqref{NESS_expansion}, we can now analyze and solve Eq.~\eqref{exact_NESS_equation} in an order-by-order manner. Let $d$ be the dimension of the Hilbert space. Thus $\genrho$ is a $d \times d$ matrix. Taking the $\epsilon\rightarrow 0$ limit of Eq.~\eqref{exact_NESS_equation} yields, in combination with Eq.~\eqref{NESS_expansion}, 
\begin{align} \label{l0_exact}
 \mathcal L_0[\genrho_{\rm NESS}^{(0)}] = [\genrho_{\rm NESS}^{(0)}, {H}_S]=0.   
\end{align}
{Here ${H}_S = \sum_{\alpha} E_\alpha |E_\alpha \rangle \langle E_\alpha |$ is a spectral decomposition in the energy eigenbasis $\{\ket{E_\alpha}\}$.}
This implies that $\genrho_{\rm NESS}^{(0)}$ is diagonal in the energy eigenbasis of the system,
\begin{align}
	\label{diagonal_condition_exact}
	& \langle E_\alpha |\genrho_{\rm NESS}^{(0)} | E_{\nu} \rangle = 0~~\forall~\alpha \neq \nu,
\end{align}
Thus,
\begin{align} \label{rerho_zero}
	{\genrho_{\rm NESS}^{(0)} = \sum_\alpha p_\alpha \ket{E_\alpha} \bra{E_\alpha}  .}
\end{align}
The next set of equations on comparing the second-order terms in $\epsilon$ (i.e. $\epsilon^2$) in Eq.~\eqref{exact_NESS_equation} is given by
\begin{equation} \label{diagonal_values}
	\bra{ E_{\alpha} } {\mathcal{L}}_2[\genrho_{\rm NESS}^{(0)}] \ket{ E_{\alpha} }=0 \quad \forall~\alpha.
\end{equation}
This fixes the diagonal values of  $\genrho_{\rm NESS}^{(0)}$, i.e, it fixes the values of $p_\alpha$. Note that there are $d$ equations and $d$ variables, {and} therefore the set of equations uniquely determine the values of the $p_\alpha$. {Similarly, at order $\epsilon^2$, from Eq.~\eqref{exact_NESS_equation}, we have for off-diagonal elements}
\begin{equation} \label{offdiagonal_rhotwo}
	i(E_\alpha -E_\nu) \bra{ E_{\alpha} } \genrho_{\rm NESS}^{(2)}\ket{ E_{\nu} }   +\bra{ E_{\alpha} } {\mathcal{L}}_2[\genrho_{\rm NESS}^{(0)}] \ket{ E_{\nu} }=0,
\end{equation}
which holds for $\forall~\alpha\neq\nu$. {This} fixes the off-diagonal values of $\genrho^{(2)}_{\rm NESS}$ given $E_\alpha \neq E_\nu$ when $\alpha \neq \nu$ (i.e. no degeneracies). Note that {since} there are $d^2-d$ equations and $d^2-d$ variables, the set of equations uniquely determine the solutions. Finally, {at order $\epsilon^4$ in Eq.~\eqref{exact_NESS_equation} for diagonal elements, }we have
\begin{equation} \label{diagonal_rhotwo}
	\bra{ E_{\alpha} } {\mathcal{L}}_2[\genrho_{\rm NESS}^{(2)}] \ket{ E_{\alpha} }+\bra{ E_{\alpha} } {\mathcal{L}}_4[\genrho_{\rm NESS}^{(0)}] \ket{ E_{\alpha} }=0,
\end{equation}
which fixes the diagonal values of $\genrho^{(2)}_{\rm NESS}$. Most often, higher-order terms are very difficult to treat. Therefore the QME only up to the leading-order term in $\epsilon$, {that is $\epsilon^2$}, is considered. \mkdev{In the next section, we motivate the microscopically derived QME up to $\epsilon^2$, which is the Redfield Equation.}

\mkdev{\section{Redfield formalism}
\label{sec1-2}
This section is a brief review of the Redfield Equation (RE) and how the NESS obtained from RE has correct leading-order populations and coherences}. To arrive at the RE, we take the composite state of all baths $ \rho_B$ {$\equiv \rerho_B^L \otimes \rerho_B^R$ to be the thermal state.
	\begin{align}\label{initial_state}
		\rho_B^i = \frac{e^{-\beta_i H_B^i}}{Z_B^i}, ~~ i=L,R.
	\end{align}
}
Without any loss of generality, we assume ${\rm Tr}_B( H_{SB}~ \rerho(0) \otimes \rerho_B)=0$~\cite{breuer_book, Rivas_2012}. The RE equation is the full expansion in Eq.~\eqref{general_TCL_QME} truncated to second order in $\epsilon$ (i.e. up to $\epsilon^2$) {and is given by}
\begin{align}
	\label{TCL2}
	\frac{\partial{\rerho}}{\partial t}= \mathcal{L}_{0}[\rerho(t)]+ \epsilon^2 \mathcal{L}_{2}[\rerho(t)].
\end{align}
Here recall from Eq.~\eqref{L0} that ${\mathcal{L}}_0[\rerho(t)] = i [\rerho(t), H_S]$ and $\mathcal{L}_2$ is given by \cite{breuer_book},
\begin{equation}
	\begin{aligned}
		\label{RE1}
		\mathcal{L}_2[\rerho(t)]= \int_0^\infty dt^\prime \operatorname{Tr}_B [ H_{SB},[H_{SB}(-t^\prime),\rerho(t) \otimes \rerho_B]]. \\
	\end{aligned}
\end{equation}
where
\begin{align}
	H_{SB}(t) = e^{i(H_S + H_B)t}H_{SB}e^{-i(H_S + H_B)t}. 
\end{align}
Even though a CPTP map gives the actual microscopic evolution [see Eq.~\eqref{tracefin}], the microscopically derived RE violates complete positivity unless in extremely special cases {such as} rapidly oscillating off-diagonal terms in the interaction picture~\cite{breuer_book}, positive semi-definite Redfield tensor~\cite{Fleming2010}, single qubit weakly interacting with bath~\cite{Whitney_2008}. Repeating the same analysis as in section~\ref{sec1-1}, we find that since only $\mathcal{L}_0$ and $\mathcal{L}_2$ show up in Eqs.~\eqref{l0_exact} to \eqref{offdiagonal_rhotwo}, the zeroth-order RE NESS is the accurate zeroth-order NESS [Eq.~\eqref{diagonal_values}], and the coherences of second-order RE NESS are the same as the accurate second-order exact NESS [Eq.~\eqref{offdiagonal_rhotwo}] (recall that for coherences, the second-order is the leading order). \mkdev{The fact that RE zeroth-order NESS is accurate is of crucial importance in our numerical results, and is highlighted again later on at the end of Sec.~\ref{sec:acccoh}}. However, note that diagonal values of $\rerho^{(2)}_{\rm NESS}$ will {deviate from the accurate second-order NESS case}, since RE does not have an analogue of Eq.~\eqref{diagonal_rhotwo} (we omitted $\mathcal{L}_4$). 


\mkdev{While RE violates a fundamental requirement in complete positivity, it leaves a question to explore if there exists any Markovian QME satisfying the fundamental conditions we want. In the next section, we explore what restrictions these fundamental requirements bring about in our Markovian QME form.}

\mkdev{\section{Requirements in a Markovian QME}
\label{sec1-3}
In this section, we consider a generic Markovian QME and highlight how exactly its superoperator form looks like and what conditions it satisfies for our given fundamental requirements.} A generic Markovian QME is given as
\begin{align}
	\label{TCL2New}
	\frac{\partial \cvxrho}{\partial t}= {\mathcal{L}}_{0}[\cvxrho(t)]+ \epsilon^2 {\mathcal{L}}^\prime_{2}[\cvxrho(t)],
\end{align}
where we replace ${\mathcal{L}}_{2}$ of RE with ${\mathcal{L}}^\prime_{2}$ of any other QME. We use $\cvxrho$ to denote the system density matrix obtained via this QME, in contrast to $\rerho$ for the RE system density matrix. Let us look at the \mkdev{conditions} one requires on this generic Markovian QME form in Eq.~\eqref{TCL2New} to obey the given fundamental requirements.

\subsection{The CPTP condition}
\label{sec:cptp}


First, we wish to preserve complete positivity and trace (CPTP map) for which one requires a Lindbladian form as given in Eq.~\eqref{Lindblad_form}. This means that ${\mathcal{L}}^\prime_{2}$ present in Eq.~\eqref{TCL2New} has to have a Lindbladian form [Eq.~\eqref{Lindblad_form}].

\subsection{Accurate populations and thermalization}
\label{sec:accpop2}
Second, one might want the Markovian QME to recover the correct leading-order populations of the NESS 
\begin{align} \label{cvxrho_pops}
\cvxrho_{\rm NESS}^{(0)} = \genrho_{\rm NESS}^{(0)}    
\end{align}
The conditions for this have been described in Ref.~\onlinecite{Tupkary_2022} and are also explained as follows: Using the same argument as in {Sec.~\ref{sec1-1}}, we can conclude that $\cvxrho^{(0)}_{\rm NESS}$ is diagonal in energy eigenbasis [see Eq.~\eqref{diagonal_condition_exact}]. We start with writing down the RE for the given setup. 
and by solving Eq.~\eqref{diagonal_values}, we get $p_\alpha$ in $\genrho_{\rm NESS}^{(0)}$ [Eq.~\eqref{rerho_zero}]. Then, in order to obtain the correct value of $\cvxrho^{(0)}_{\rm NESS}$ as in Eq.~\eqref{cvxrho_pops}, $\mathcal{L}_2^\prime$ in Eq.~\eqref{TCL2New} must satisfy the same Eq.~\eqref{diagonal_values} as $\mathcal{L}_2$ in RE [Eq.~\eqref{TCL2}]. Thus, we must have
\begin{equation} \label{diagonal_values_prime}
	\bra{ E_{\alpha}}  {\mathcal{L}}^\prime_2[\genrho_{\rm NESS}^{(0)}] \ket{ E_{\alpha} }= \bra{ E_{\alpha} } {\mathcal{L}}_2[\genrho_{\rm NESS}^{(0)}] \ket{ E_{\alpha} }=0 ,~~\forall~\alpha.
\end{equation}
In the equilibrium ($\beta_L = \beta_R = \beta$) scenario, Eq.~\eqref{diagonal_values_prime} becomes the condition for thermalization~\cite{Tupkary_2022}. Thermalization implies that after an infinite time, the steady state reaches the thermal state at a temperature defined by $\beta = 1/T$ when all the baths attached to the system are at temperature $T$, or in other words, we have $p_\alpha \propto e^{-\beta E_\alpha}$ in Eq.~\eqref{diagonal_values_prime}. The equilibrium case is just a special case of the non-equilibrium ($\beta_L \neq \beta_R$) scenario where one has to solve the linear equations obtained from RE to obtain $p_\alpha$.

\subsection{Accurate coherences}
\label{sec:acccoh2}
Third, to recover the leading-order coherences of the NESS, i.e. 
\begin{align}
 \bra{E_\alpha}\cvxrho_{\rm NESS}^{(2)}\ket{E_\nu} = \bra{E_\alpha}\genrho_{\rm NESS}^{(2)}\ket{E_\nu},~\forall~\alpha \neq \nu 
\end{align}
${\mathcal{L}}_2^\prime$  must yield exactly the same set of equations as Eq.~\eqref{offdiagonal_rhotwo}. Therefore, 
\mkdev{\begin{equation} \label{offdiagonal_condition_og}
	\bra{ E_{\alpha} } {\mathcal{L}}^\prime_2[\bar \genrho_{\rm NESS}^{(0)}] \ket{ E_{\nu} } =  \bra{ E_{\alpha} } {\mathcal{L}}_2[\genrho_{\rm NESS}^{(0)}] \ket{ E_{\nu}  } ~~ \forall~\alpha \neq \nu.
\end{equation}
}
Given that one has obtained $\cvxrho^{(0)}_{\rm NESS} = \genrho_{\rm NESS}^{(0)}$, we can write
\begin{equation} \label{offdiagonal_condition}
	\bra{ E_{\alpha} } {\mathcal{L}}^\prime_2[\genrho_{\rm NESS}^{(0)}] \ket{ E_{\nu} } =  \bra{ E_{\alpha} } {\mathcal{L}}_2[\genrho_{\rm NESS}^{(0)}] \ket{ E_{\nu}  } ~~ \forall~\alpha \neq \nu.
\end{equation}

\subsection{Local conservation laws}
\label{sec:locc}
Finally, we require local conservation laws. For this, consider the dynamical equation for the expectation value of any observable $O_M$ on $\mathcal{H}_M$ [see Fig.~\ref{fig:main-img}], which is given by
\begin{align} \label{loccons}
	\frac{d}{dt} \langle I_L \otimes O_M \otimes I_R \rangle = -i \langle [I_L \otimes O_M \otimes I_R, H_S] \rangle, 
\end{align}
where $\langle X \rangle = \text{Tr}[X\rho]$. \mkdev{This equation is arrived at by the simple physical idea that time evolution of any system operator $O$ in the Heisenberg picture, which commutes with the system bath operator, is given as
\begin{align} \label{eq:loc_idea}
   \frac{d \langle O \rangle}{dt} &= -i \langle [O,H_S]\rangle,~\forall ~[O,H_{SB}] = 0. 
\end{align}
Given an energy eigenbasis decomposition of $H_S = \sum_a E_a \ket{E_a}\bra{E_a}$, we can write Eq.~\eqref{eq:loc_idea} as
\begin{align} \label{eq:loc_currents}
   \frac{d \langle O \rangle}{dt} &= -i \sum_{a,\nu=1}^D (E_a-E_\nu) \langle E_a|\rho|E_\nu \rangle\langle E_\nu | O|E_a\rangle \notag\\
   &\forall ~[O,H_{SB}] = 0. 
\end{align}
We see that the form of our system operator in Eq.~\eqref{loccons} commuted with the system-bath coupling Hamiltonian as required for equality in Eq.~\eqref{eq:loc_idea}.  Our preservation of local conservation laws here is more general.} Any effective QME obtained by integrating out the bath should satisfy this property. We call QMEs satisfying this property [Eq.~\eqref{loccons}] as ones preserving local conservation laws. The most general form of Markovian QME respecting the restrictions for satisfying complete positivity [Eq.~\eqref{Lindblad_form}] \textit{and} local conservation laws [Eq.~\eqref{loccons}] was shown in Ref.~\onlinecite{tupkary_searching_2023} to be one where the Lindblad operators and the Lamb Shift Hamiltonian only act on the part of the system connected to the bath, and are identity elsewhere. Thus, it is given by


%
\begin{widetext}
	\begin{equation}
		\begin{aligned} \label{eq:local_lindblad_for_TOP}
			\frac{\partial {\cvxrho}}{\partial t} &= i [\cvxrho,H_S+\epsilon^2 H^{(L)}_{LS} \otimes I_{MR} +  \epsilon^2  I_{LM} \otimes H^{(R)}_{LS}]\\
			&+ \epsilon^2 \sum_{\alpha_L, {\tilde{\alpha}}_L=1}^{d_L^2-1} \Gamma^{(L)}_{\alpha_L, {\tilde{\alpha}}_L}  
			\bigg( \left(f_{\tilde{\alpha}_L} \otimes \frac{I_{MR}}{\sqrt{d_Md_R}}\right) \cvxrho \left(f_{\alpha_L} \otimes \frac{I_{MR}}{\sqrt{d_Md_R}}\right)^\dagger
			- \frac{\{ (f_{\alpha_L} \otimes I_{MR})^\dagger (f_{\tilde{\alpha}_L} \otimes I_{MR}),\cvxrho \}}{2d_Md_R}  \bigg) \\
			&+ \epsilon^2 \sum_{\alpha_R, {\tilde{\alpha}}_R=1}^{d_R^2-1} \Gamma^{(R)}_{\alpha_R, {\tilde{\alpha}}_R}  
			\bigg( \left(\frac{I_{LM}}{\sqrt{d_Ld_M}} \otimes f_{\tilde{\alpha}_R}\right) \cvxrho \left(\frac{I_{LM}}{\sqrt{d_Ld_M}} \otimes f_{\alpha_R}\right)^\dagger
			- \frac{\{ (I_{LM} \otimes f_{\alpha_R})^\dagger (I_{LM} \otimes f_{\tilde{\alpha}_R}),\cvxrho \}}{2d_Ld_M}  \bigg),
		\end{aligned}
	\end{equation}
\end{widetext}
where $I_{MR} = I_M \otimes I_R,~I_{LM} = I_L \otimes I_M$, where $I_L$ and $I_R$ are identity operators on the Hilbert spaces $\mathcal{H}_L$ and $\mathcal{H}_R$ [see Fig.~\ref{fig:main-img}]. Here $d_L~(d_R)$ is the dimension of the Hilbert space $\mathcal H_L~(\mathcal{H}_R)$. $d = d_L d_M d_R$ is the dimension of the total Hilbert space $\mathcal H = \mathcal H_L \otimes \mathcal H_M \otimes \mathcal{H}_R$, where $d_M$ is the dimension of the Hibert space $\mathcal H_M$. $\Gamma^{(L)}_{\alpha_L, {\tilde{\alpha}}_L},~ \Gamma^{(R)}_{\alpha_R, {\tilde{\alpha}}_R}$ are positive-semidefinite matrices, $H_{LS}^{(L)},~H_{LS}^{(R)}$ are Hermitian and $f_{\alpha_L},~f_{\alpha_R}$ are orthonormal basis of operators in the parts of the system connected to the left and right baths respectively, where orthonormality is defined according to the Hilbert Schmidt
inner product given by $\braket{A}{B}= \Tr(A^\dagger B)$. We choose the basis such that $f_{\alpha_L = d_L^2} = I_L/\sqrt{d_L},~f_{\alpha_R = d_R^2} = I_R/\sqrt{d_R}$.

\mkdev{Here, we see from Eq.~\eqref{eq:loc_currents} that up to leading-order in $\epsilon$ for a QME, even if one has preservation of local conservation laws [as in Eq.~\eqref{loccons}], whether or not the time evolution of the expectation value of the operator is correct (i.e. matches the RE) or not depends on whether the QME state gives the correct leading-order coherences. For the state $\cvxrho$ in generic Markovian QME given in Eq.~\eqref{TCL2}, we can have the case that it preserves local conservation laws but gives inaccurate leading-order coherences, which implies that for all $[O,H_{SB}] = 0$, up to leading-order 
\begin{align}
    \frac{d \langle O \rangle_{\genrho}}{dt} &= -i \langle [O,H_S] \rangle_{\genrho} \neq -i \langle [O,H_S] \rangle_{\cvxrho} = \frac{d \langle O \rangle_{\cvxrho}}{dt},
\end{align}
with $\langle A \rangle_\rho = \Tr(A\rho)$. For the steady-states, it would mean that for the generic QME NESS $\cvxrho_{\rm NESS}$, while the non-equilibrium currents moving into and out of the bulk would be equal (satisfying local energy conservation in the case of energy currents for example), they would be inaccurate. While this may prompt one to think of a QME giving correct leading-order coherences but violating local conservation laws, we show that this case is not possible (up to leading-order) in Appendix~\ref{appn4}.}

\mkdev{More importantly, we wish to highlight how preservation of local conservation laws does not imply accurate leading-order coherences (and equivalently currents for NESSs). Since we take an LE form as in Eq.~\eqref{eq:local_lindblad_for_TOP} satisfying CPTP and local conservation, it remains to be seen whether any such form can also obtain NESSs with correct leading-order coherences.}


These [Secs.~\ref{sec:cptp}, \ref{sec:locc}, \ref{sec:accpop2}, \ref{sec:acccoh2}] are the four fundamental requirements we have considered in this work. \mkdev{We remark that except complete positivity, RE satisfies all the other fundamental requirements as shown in Ref.~\onlinecite{Tupkary_2022}. In the next section, we show how to search for Markovian QMEs satisfying all these requirements.}
\mkdev{\section{Semidefinite programming formulation of fundamental requirements} 
\label{sec1-4}
In this section, we show how to formulate optimizable metrics on accurate populations and coherences given a Markovian QME form which satisfies CPTP and local conservation laws (i.e. given an LE form, see Secs.~\ref{sec:cptp},\ref{sec:locc}), which comes under the the semidefinite programming paradigm.}

\subsection{{CPTP and} Local conservation laws with accurate populations}
\label{sec:accpop}

Let us investigate the conditions necessary for obtaining a Markovian QME in Lindlbad form {(or LE)} [{Sec.~\ref{sec:cptp}}] that satisfies local conservation laws [{Sec.~\ref{sec:locc}}] and yields accurate populations [{Sec.~\ref{sec:accpop2}}].

To ensure local conservation laws hold and the QME is in Lindblad form, we ensure that our ${\mathcal{L}}^\prime_2[\genrho_{\rm NESS}^{(0)}]$ is of the form given in Eq.~\eqref{eq:local_lindblad_for_TOP}. If one is additionally interested in checking if Eq.~\eqref{diagonal_values_prime} is satisfied or not, one can simply check whether the following quantity $\tau^{\rm pop}$ equals zero or not, where
\begin{equation} \label{tauequation}
	\tau^{\rm pop} = \sum_\alpha \lvert  \bra{ E_{\alpha} } {\mathcal{L}}^\prime_2[\genrho_{\rm NESS}^{(0)}] \ket{ E_{\alpha} } \rvert.
\end{equation}
$\tau^{\rm pop}$ is a function of the fixed value of $\genrho_{\rm NESS}^{(0)}$, the zeroth-order NESS, while the variable to be optimized over  is ${\mathcal{L}}^\prime_2$ consisting of $H_{LS}^{(i)},~\Gamma^{(i)}$, $i \in \{L,R\}$.
Note that $\tau^{\rm pop} = 0$ iff Eq.~\eqref{diagonal_values_prime} holds, making it sufficient to work with $\tau^{\rm pop}$ to see if Eq.~\eqref{diagonal_values_prime} is satisfied. This can be formulated in the form of an optimization problem, in the same manner as done for the thermalization optimization problem (TOP) in Ref.~\onlinecite{tupkary_searching_2023}. Here we consider the more general $\beta_L \neq \beta_R$ non-equilibrium case rather than the equilibrium scenario in the TOP, as given by varying $H_{LS},~\Gamma$ for both left and right baths. We remind the reader that we only require to solve $d$ equations in $d$ variables to obtain $p_\alpha$ in Eq.~\eqref{rerho_zero} for this optimization problem in the non-equilibrium scenario.
\begin{align}\label{eq:diag_optimization}
	\text{minimize : } \quad &\tau^{\rm pop} ~ \textrm{by varying } H^{(L)}_{LS},~H^{(R)}_{LS},~\Gamma^{(L)},~\Gamma^{(R)} \notag \\
	\text{subject to : } \quad & H^{(L)}_{LS},~H^{(R)}_{LS} \text{ are hermitian,} \\
	& \text{Tr}(\Gamma^{(L)}) = 1,~~ \Gamma^{(L)} \geq 0, \notag \\
	& \text{Tr}(\Gamma^{(R)}) = 1,~~ \Gamma^{(R)} \geq 0, \notag
\end{align}
Here $\Gamma^{(i)}\geq 0$ means $\Gamma^{(i)}$ is a positive semi-definite matrix. Note that the choice of $\text{Tr}(\Gamma^{(L)}) = \text{Tr}(\Gamma^{(R)}) = 1$ in Eq.~\eqref{eq:diag_optimization} is completely arbitrary scaling, and can be set to other fixed values as well. In the optimization problem, let $\tau^{\rm pop}_{\rm opt}$ be the optimal value obtained from solving Eq.~\eqref{eq:diag_optimization}. Given a tolerance $\delta_{\rm tol}$,
\begin{align}
	\label{eq:tolerance}
	\textrm{if } \begin{cases} \tau^{\rm pop}_{\rm opt} <\delta_{\rm tol}, &\textrm{ the desired LE is possible}, \\
		\tau^{\rm pop}_{\rm opt} \geq \delta_{\rm tol} & \textrm{ the desired LE is impossible}, \end{cases}
\end{align}
For $\tau^{\rm pop}_{\rm opt} > 0$, it will be interesting to know how far away the zeroth-order non-equilibrium steady state ($\cvxrho_{\rm NESS}^{(0)}$) obtained via any LE satisfying local conservation laws [Eq.~\eqref{eq:local_lindblad_for_TOP}] is from the zeroth-order NESS obtained from RE ($\rho_{\rm NESS}^{(0)}$), as a way of understanding how ``good" the leading-order NESS obtained from the LE is going to be. This can be calculated in the form of trace distance between the two density matrices, which we present in the form of a lemma. 

\begin{restatable}{lm}{tracedistlem} \label{lm:trace_dist}
	Let $\cvxrho_{\rm NESS}^{(0)}$ be the zeroth-order NESS obtained via any LE satisfying local conservation laws [Eq.~\eqref{eq:local_lindblad_for_TOP}], and $\genrho^{(0)}_{\rm NESS}$ as the zeroth-order NESS. Given that $\text{Tr}(\Gamma^{(i)}) = t_i,~i\in \{L,R\}$, then
	\begin{widetext}
		\begin{align}
			\tracedist((\cvxrho^{(0)}_{\rm NESS}, \rho^{(0)}_{\rm NESS}) &\geq \frac{\tau^{\rm pop}_{\rm opt}}{2(t_Ld^{{3}}_L +t_Rd^{{3}}_R - t_L{{d_L}}-t_R{{d_R}})} \label{lower_bound_maintxt}
		\end{align}
	\end{widetext}
	where $\tau^{\rm pop}_{\rm opt}$ is the optimal value obtained from Eq.~\eqref{eq:diag_optimization}, $d$ is the dimension of the Hilbert space of the system, while $d_L~(d_R)$ is the dimension of the Hilbert space of the part of the system connected to the left (right) bath [see Fig.~\ref{fig:main-img}].
\end{restatable}
Details of the proof of this Lemma is given in appendix~\ref{appn2}. This result highlights that for any non-zero $\tau^{\rm pop}_{\rm opt}$ value, the trace distance between the zeroth-order NESS and the NESS obtained from \textit{any} LE satisfying local conservation laws has a non-zero lower bound of $\tau^{\rm pop}_{\rm opt}/\alpha$, {where, from Eq.~\eqref{lower_bound_maintxt}, $\alpha = 2\left(t_L(d_L^{{3}}-{d_L}) +t_R(d^{{3}}_R-{d_R})\right)$}. 



\subsection{{CPTP and} Local conservation laws with accurate populations and coherences} \label{sec:acccoh}

Just as in the last subsection, if one is interested in checking whether both Eqs.~\eqref{diagonal_values_prime} and \eqref{offdiagonal_condition} are satisfied or not (along with local conservation laws), then it is equivalent to checking whether ${\mathcal{L}}^\prime_2[\genrho_{\rm NESS}^{(0)}]$ equals ${\mathcal{L}}_2[\genrho_{\rm NESS}^{(0)}]$. Note that ${\mathcal{L}}_2[\genrho_{\rm NESS}^{(0)}]$ is guaranteed to be zero along the diagonal by the construction of $\genrho_{\rm NESS}^{(0)}$. Thus, we can check whether $\tau^{\rm pop,coh}$ equals zero or not, where $\tau^{\rm pop,coh}$ is defined as 
\begin{equation} \label{taucohequation}
	\tau^{\rm pop,coh} = \norm{{\mathcal{L}}^\prime_2[\genrho_{\rm NESS}^{(0)}] - {\mathcal{L}}_2[\genrho_{\rm NESS}^{(0)}] }_2.
\end{equation}
Eq.~\eqref{taucohequation} highlights how far ${\mathcal{L}}^\prime_2[\genrho_{\rm NESS}^{(0)}]$ is from ${\mathcal{L}}_2[\genrho_{\rm NESS}^{(0)}]$, in the form of the $2-\text{\rm norm}$ function, although this choice of norm is arbitrary and any $p-\text{norm}$ ($||\cdots||_p$) can be used (fundamentally, we only require that the resulting optimization problem be an SDP).  When ${\mathcal{L}}^\prime_2[\genrho_{\rm NESS}^{(0)}] = {\mathcal{L}}_2[\genrho_{\rm NESS}^{(0)}]$, $\tau^{\rm pop,coh} = 0$, and we have a LE whose NESS recovers correct populations and coherences while satisfying local conservation laws. This can again be formulated as an optimization problem given by
\begin{align}\label{eq:coh_optimization}
	&\text{minimize : }  \tau^{\rm pop,coh} ~ \textrm{by varying } H^{(L)}_{LS},~H^{(R)}_{LS},~\Gamma^{(L)},~\Gamma^{(R)} \notag \\
	&\text{subject to : }  H^{(L)}_{LS},~H^{(R)}_{LS} \text{ are hermitian,} \\
	& ~~~~~~~~~~~~~~~~\text{Tr}(\Gamma^{(L)}) = 1,~~ \Gamma^{(L)} \geq 0, \notag \\
	& ~~~~~~~~~~~~~~~~\text{Tr}(\Gamma^{(R)}) = 1,~~ \Gamma^{(R)} \geq 0, \notag
\end{align}
Similarly {as in Sec.~\ref{sec:accpop}}, let $\tau^{\rm pop,coh}_{\rm opt}$ be the optimal value obtained from solving Eq.~\eqref{eq:coh_optimization}. With the same tolerance $\delta_{\rm tol}:$
\begin{align}
	\label{eq:tolerance_coh}
	\textrm{if } \begin{cases} \tau^{\rm pop,coh}_{\rm opt} <\delta_{\rm tol}, &\textrm{ desired LE is possible}, \\
		\tau^{\rm pop,coh}_{\rm opt} \geq \delta_{\rm tol} & \textrm{ desired LE is impossible}, \end{cases}
\end{align}
Notice that both Eqs.~\eqref{eq:diag_optimization} and \eqref{eq:coh_optimization} are SDPs, and are similar to the TOP problem in Ref.~\onlinecite{tupkary_searching_2023}. Therefore, they can be directly put into standard packages for disciplined convex optimization like the CVX MATLAB~\cite{cvx} package which automatically outputs the correct optimal value and gives one set of {optimal} values for $H^{(L)}_{LS},~H^{(R)}_{LS},~\Gamma^{(L)}$ and $\Gamma^{(R)}$. Thus, if $\tau^{\rm pop}_{\rm opt}, \tau^{\rm pop,coh}_{\rm opt} < \delta_{\rm tol}$, it not only says that the desired type of LE \mkdev{is possible (up to our given tolerance)} but it also outputs one possible candidate for such a LE. If $\tau^{\rm pop}_{\rm opt}, \tau^{\rm pop,coh}_{\rm opt} \geq \delta_{\rm tol}$, the desired type of LE is impossible. Notice that satisfying $\tau^{\rm pop,coh}_{\rm opt} \leq \delta_{\rm tol}$ implies that getting correct coherences of second-order NESS and correct populations of zeroth-order NESS may be possible via an LE, but $\tau^{\rm pop,coh}_{\rm opt} > \delta_{\rm tol}$ implies that either the LE NESS fails to accurately capture the correct leading-order populations and/or accurately capture the correct leading-order coherences. In other words, for $\tau^{\rm pop,coh}_{\rm opt} > \delta_{\rm tol}$, there exists no LE whose NESS gets the correct leading-order populations and coherences.

We {again} remark that in all of the above, $p_\alpha$ [Eq.~\eqref{rerho_zero}] can be easily obtained by solving the RE. Moreover, one only has to solve $d$ equations in $d$ variables to obtain $p_\alpha$ from RE. This task is much easier than solving the RE itself. Also note here that checking if LE obtains accurate leading-order coherences separately from accurate leading-order populations [Eq.~\eqref{offdiagonal_rhotwo} with $\cvxrho^{(0)}_{\rm NESS}$ instead of $\genrho^{(0)}_{\rm NESS}$ in the second term] is not an SDP. This is because we require $\cvxrho^{(0)}_{\rm NESS}$ to be computed before that, which means $\mathcal L_2 ^\prime$ and $\cvxrho^{(0)}_{\rm NESS}$ are both variables in this case only satisfying Eq.~\eqref{diagonal_values} (here $\cvxrho^{(0)}_{\rm NESS}$ in general need not be equal to $\genrho^{(0)}_{\rm NESS}$). 

\section{Discussion of {Numerical} results}
\label{sec:sec3}

\begin{table*}[t]
	\centering
	\begin{tabular}{|c|c|c|c|c|c|c|c|c|}
		\hline
		$N_L = N_R$   & $\tau^{\rm pop}_{\rm opt}$ v/s $\beta$ &$\tau^{\rm pop,coh}_{\rm opt}$ v/s $\beta$ & $\tau^{\rm pop}_{\rm opt}$ v/s $g$ & $\tau^{\rm pop,coh}_{\rm opt}$ v/s $g$ & Correct populations & Correct coherences & Local conservation laws\\ \hline \hline
		1   & Fig.~\ref{fig:Nl1}(a) & Fig.~\ref{fig:Nl1}(b) & Fig.~\ref{fig:Nl1}(c)& Fig.~\ref{fig:Nl1}(d) & Impossible & Impossible & Satisfied\\ \hline
		
		2   & Fig.~\ref{fig:Nl2}(a) & Fig.~\ref{fig:Nl2}(b) & Fig.~\ref{fig:Nl2}(c) & Fig.~\ref{fig:Nl2}(d) & Possible & Impossible & Satisfied\\ \hline
		
	\end{tabular}
	\caption{\label{Table:figs_main} Table of figure references {summarising our results in the non-equilibrium ($\beta_L \neq \beta_R$) scenario} for different energy biases and number of qubits $N_L,~N_R$ attached to the left and right baths {respectively}. All plots are done for system-bath coupling $\epsilon = 0.01$. The columns for ``Correct populations" and ``Correct coherences" highlight the possibility or impossibility for obtaining correct {leading-order} diagonal and off-diagonal terms {respectively in} the {steady state in a wide parameter regime}, while still maintaining weak system-bath coupling.   In {very special scenarios like in }equilibrium and extremely low inter-site coupling $g$, it might be {feasible} to get correct populations and/or correct coherences.}
\end{table*}

We will now use the formulated optimization problems to analyze a specific system-bath setup. By implementing our semidefinite programming framework, we aim to determine whether a physically consistent {LE} can be constructed for this setup while satisfying the desired properties laid down in Sec.~\ref{sec1-3}. This analysis will provide insights into the feasibility of different dynamical behaviors and potentially reveal fundamental limitations on the existence of such LEs.

In particular, we study the possibility of having a Lindblad description satisfying local conservation laws [Sec.~\ref{sec:locc}] and obtaining correct populations [Secs.~\ref{sec:accpop2},~\ref{sec:accpop}] and coherences [Secs.~\ref{sec:acccoh2},~\ref{sec:acccoh}] of the zeroth-order and second-order, for non-equilibrium (or equilibrium with $\beta_L = \beta_R$) steady states in open isotropic XXZ qubit chain system with some of the qubits attached to baths {modelled by an infinite number of bosonic modes}. We will start by describing our setup. The Hamiltonian for this setup is given by $H=H_S + H_{SB}+H_B$ with
\begin{align*}
 H_S &= \omega_0 \sum_{\ell=1}^N {\sigma}_z^{(\ell)}  +  g \sum_ {\ell=1}^{N-1} \big( {\sigma}_x^{(\ell)} {\sigma}_x^{(\ell+1)} + {\sigma}_y^{(\ell)} {\sigma}_y^{(\ell+1)} \\
		&+  {\sigma}_z^{(\ell)} {\sigma}_z^{(\ell+1)} \big),   
\end{align*}
\begin{equation}  \label{eq:ham}
	\begin{aligned}
		& {H}_{SB}  = \sum_{\ell \in \{N_L, N_R\}}\sum_{r=1}^\infty  (\kappa_{\ell r} {B}^{ {(\ell)} \dagger}_r {\sigma}^{(\ell)}_- + \kappa_{lr}^* {B}^{(\ell)}_r {\sigma}^{(\ell)}_+), \\
		&{H}_B= \sum_{\ell \in \{N_L, N_R\}} \sum_{r=1}^\infty \Omega^\ell_r {B}^{{(\ell)} \dagger}_r {B}^{(\ell)}_r,
	\end{aligned}
\end{equation}
where $\sigma^{(\ell)}_{x,y,z}$ denotes the Pauli matrices acting on the $\ell^{\text{th}}$ qubit, ${\sigma}^{(\ell)}_{+}=({\sigma}^{(\ell)}_{x}+i {\sigma}^{(\ell)}_{y})/2$, ${\sigma}^{(\ell)}_{-}=({\sigma}^{(\ell)}_{x}-i {\sigma}^{(\ell)}_{y})/2$, ${B}^{(\ell)}_r$ is bosonic annihilation operator for the $r^{\text{th}}$ mode of the bath attached at the $\ell^{\text{th}}$ site. Here $\omega^{(\ell)}_0$ and $g$ represent the magnetic field and the overall qubit-qubit coupling strength. The first $N_L$ and the last $N_R$ qubits are attached to the left and right baths, while the remaining $N_M = N - N_L -N_R$ qubits are not attached to any bath. {The specific setup falls under the schematic structure of Fig.~\ref{fig:main-img}}. \mkdev{Note that the condition listed in Eq.~\eqref{eq:odd_point} becomes a condition for odd-point bath correlators to be zero in our setup, which is true for our Gaussian bosonic baths due to Wick's theorem~\cite{Wick_1950}}.

\begin{figure*}
	\centering
	\hspace*{-10mm}
	\includegraphics[width=2\columnwidth]{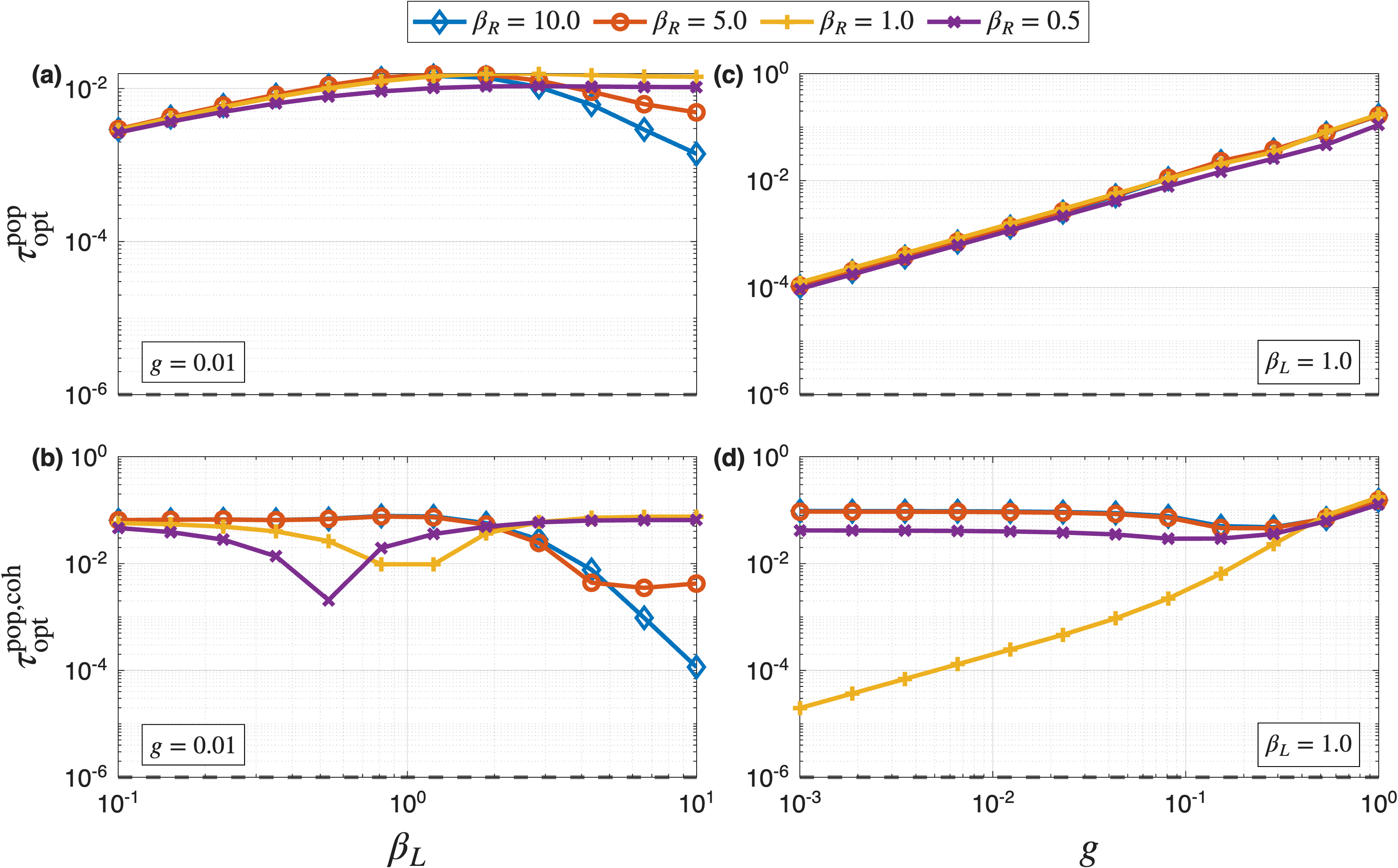}
	\caption{Plots for $N_L = N_R = 1,~N_M = 2$ for the isotropic XXZ Hamiltonian [Eq.~\eqref{eq:ham}] keeping $\gamma_\ell = 1~ \forall \ell,~\omega_c=10$ [Eq.~\eqref{eq:ohmic_bath}] and $\beta_R = 10.0$ (blue diamond), $5.0$ (orange circle), $1.0$ (yellow plus), $0.5$ (purple cross). {The black dashed horizontal line represents {the chosen tolerance} $\delta_{\rm tol} = 10^{-6}$}. (a) $\tau^{\rm pop}_{\rm opt}$ [Eq.~\eqref{eq:diag_optimization}] versus $\beta_L$ with $g=0.01$. (b) $\tau^{\rm pop,coh}_{\rm opt}$ [Eq.~\eqref{eq:coh_optimization}] versus $\beta_L$ with $g=0.01$. (c) $\tau^{\rm pop}_{\rm opt}$ versus $g$ for $\beta_L = 1.0$. (d) $\tau^{\rm pop,coh}_{\rm opt}$ versus $g$ for $\beta_L = 1.0$. This figure shows that it is impossible to achieve populations correctly up to leading-order (implying incorrect \mkdev{populations and} coherences \mkdev{together} as well)  for only single qubits attached to left and right baths as discussed in Sec.~\ref{sec:nl1}. \mkdev{One can also see the impossibility of achieving correct leading-order populations as in panels (a,c) as a consequence of Lemma~\ref{lm:trace_dist}, which gives an above tolerance non-zero lower bound between the CVX obtained and the RE (i.e. accurate) zeroth-order NESSs.}}
	\label{fig:Nl1}
\end{figure*}


\begin{figure*}
	\centering
	\hspace*{-10mm}
	\includegraphics[width=2\columnwidth]{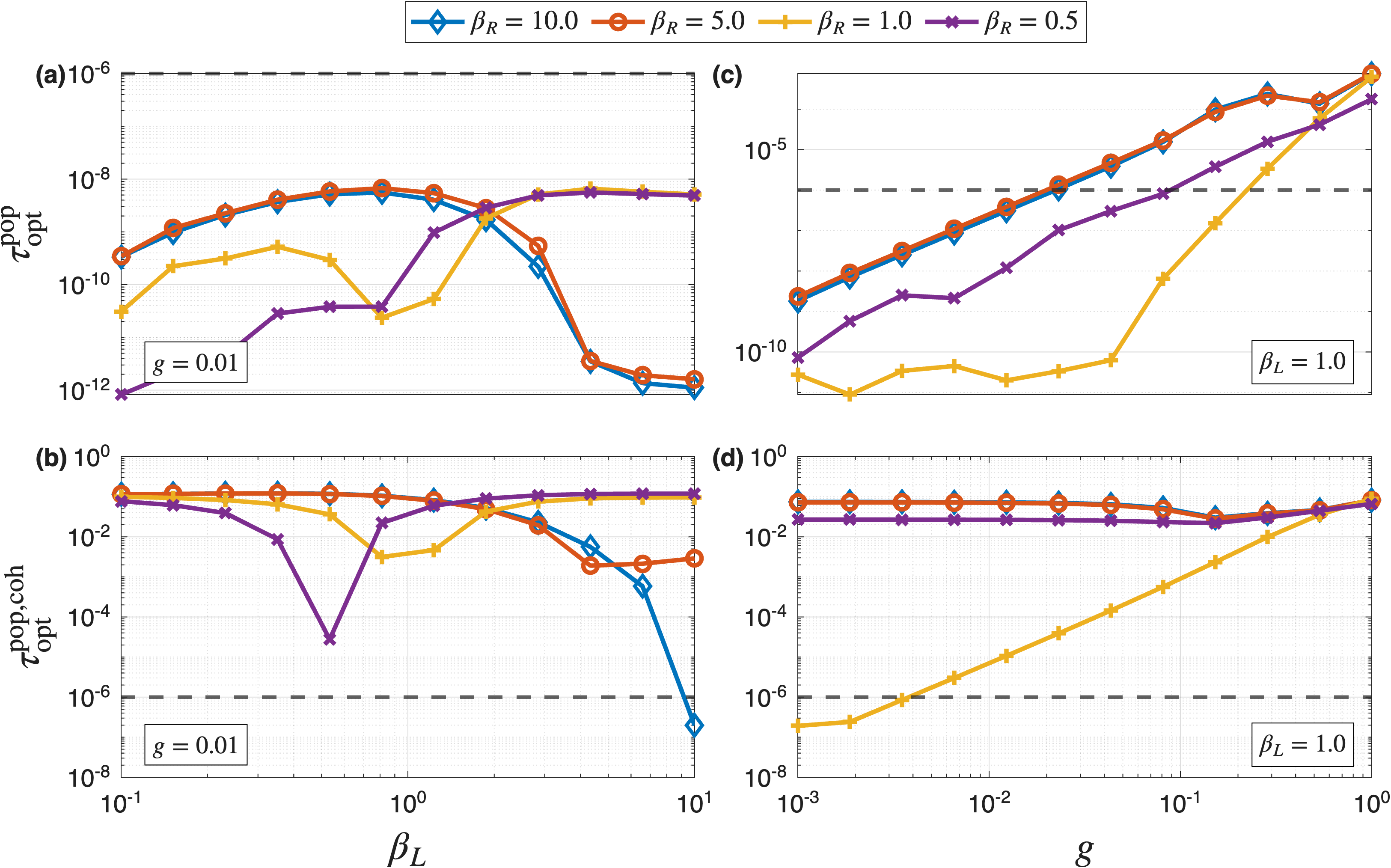}
	\caption{Plots for $N_L = N_R = 2,~N_M = 2$ ({i.e. two qubits attached to left and right baths, see Sec.~\ref{sec:nl2})} for the isotropic XXZ Hamiltonian [Eq.~\eqref{eq:ham}] keeping $\gamma_\ell = 1~ \forall \ell,~\omega_c=10$ [Eq.~\eqref{eq:ohmic_bath}] and $\beta_R = 10.0$ (blue diamond), $5.0$ (orange circle), $1.0$ (yellow cross), $0.5$ (purple plus). {The black dashed horizontal line represents $\delta_{\rm tol} = 10^{-6}$}. (a) $\tau^{\rm pop}_{\rm opt}$ [Eq.~\eqref{eq:diag_optimization}] versus $\beta_L$ with $g=0.01$. (b) $\tau^{\rm pop,coh}_{\rm opt}$ [Eq.~\eqref{eq:coh_optimization}] versus $\beta_L$ with $g=0.01$. (c) $\tau^{\rm pop}_{\rm opt}$ versus $g$ for $\beta_L = 1.0$. (d) $\tau^{\rm pop,coh}_{\rm opt}$ versus $g$ for $\beta_L = 1.0$. This figure shows that it is possible (up to $\delta_{\rm tol}$) to obtain steady-states with correct leading-order populations for low inter-site coupling, but it is impossible to obtain both correct leading-order populations and coherences generally.}
	\label{fig:Nl2}
\end{figure*}

{At initial time, the baths are taken to be in their respective thermal state with inverse temperatures $\beta_\ell$ and chemical potentials $\mu_\ell$.  The dynamics of the system can be shown to be governed by the bath spectral functions, defined as
	\begin{align}
		\mathfrak{J}_\ell(\omega)=  2 \pi \sum_{k=0}^\infty  \left| \kappa_{\ell k} \right|^2 \delta (\omega-\Omega^\ell_k) \label{Jw}
	\end{align}
	and the Bose distribution,
	\begin{align}
		n_\ell(\omega)=[e^{\beta_\ell\omega} - 1]^{-1},
	\end{align}
	   corresponding to the initial states of the baths. The RE for this set-up is obtained by simplification of Eq.~(\ref{RE1}), which gives the following.   
	\begin{align}
		\label{RE2}
		\frac{\partial{\rho}}{\partial t} &= i[{\rho},{H}_S] \nonumber  \\
		&- \epsilon^2 \sum_\ell \Big ( \big[{S}_\ell^\dagger, {S}_\ell^{(1)} {\rho}\big] + \big[{\rho} {S}_\ell^{(2)}, {S}_\ell^\dagger \big] + {\rm H.c.} \Big),
	\end{align}
	where $S_\ell$ is the system operator at $\ell$th site coupling with the corresponding bath and
	\begin{align}
		&{S}_\ell^{(1)}=\int_0^\infty dt^\prime \int\frac{d\omega}{2\pi}{S}_\ell(-t^\prime) e^{\beta_\ell(\omega-\mu_\ell)}\mathfrak{J}_\ell(\omega)n_\ell(\omega)e^{-i\omega t^\prime}, \nonumber \\  
		&{S}_\ell^{(2)}=\int_0^\infty dt^\prime \int\frac{d\omega}{2\pi}{S}_\ell(-t^\prime)  \mathfrak{J}_\ell(\omega)n_\ell(\omega)e^{-i\omega t^\prime}, 
	\end{align}
	with ${S}_\ell(t)=e^{i{H}_S t} {S}_\ell e^{-i{H}_S t}$, and H.c. denoting Hermitian conjugate. For our setting, we have ${S}_\ell = \sigma_-^{(\ell)}$. Putting this in Eq.~\eqref{RE2}, we note that the second-order dissipator (${\mathcal{L}}_2$) of the RE is \cite{Tupkary_2022}
	\begin{align} \label{l2_red_maintxt}
		{\mathcal{L}}_2(\rho) &= \sum_\ell  \sum_{\alpha, \gamma = 1}^{2^N}\{[\rho \proj{E_\alpha} \sigma_-^{(\ell)} \proj{E_\gamma},\sigma_+^{(\ell)}]C_\ell(\alpha,\gamma) \notag\\
		&+ [\sigma_+^{(\ell)}, \proj{E_\alpha}\sigma_-^{(\ell)}\proj{E_\gamma}]D_\ell(\alpha,\gamma) + \text{H.c.}\},
	\end{align}
	with
	\begin{equation}
		\begin{aligned}
			C_{\ell}(\alpha,\gamma) &= \frac{\mathfrak{J}_{\ell}(E_{\gamma \alpha}) n_{\ell}(E_{\gamma \alpha})}{2 } - i \mathcal{P} \int_{0}^{\infty} d \omega \frac{\mathfrak{J}_{\ell}(\omega) n_{\ell}(\omega)}{\omega-E_{\gamma \alpha}},  \\
			D_{\ell}(\alpha,\gamma) &= \frac{ e^{\beta_{\ell}(E_{\gamma \alpha}-\mu_{\ell})} \mathfrak{J}_{\ell}(E_{\gamma \alpha}) n_{\ell}(E_{\gamma \alpha})}{2 }\\ &- i \mathcal{P} \int_{0}^{\infty} d \omega \frac{e^{\beta_{\ell}(\omega-\mu_{\ell})} \mathfrak{J}_{\ell}(\omega) n_{\ell}(\omega)}{\omega-E_{\gamma \alpha}},
			\label{redfield:constants}
		\end{aligned}
	\end{equation}
	where $\mathcal{P}$ denotes the Cauchy Principal value, and $E_{\gamma \alpha} = E_\gamma - E_\alpha$. We consider bosonic baths described by Ohmic spectral functions with Gaussian cut-offs, 
	\begin{align} \label{eq:ohmic_bath}
		\mathfrak{J}_\ell(\omega)=\gamma_\ell\omega e^{-(\omega / \omega_c)^2}\Theta(\omega),
	\end{align}
	 where, $\Theta(\omega)$ is Heaviside step function, $\gamma_\ell$ is the coupling strength between $\ell^{\rm th}$ qubit and bath and $\omega_c$ is the cut-off frequency (set to $\omega_c=10$ for all figures in this work).}

We are interested in finding a Markovian description that accurately captures the steady-state density matrix in the above setting.
As mentioned before in Sec.~\ref{sec1-2}, the RE is known to give accurate populations and coherences to leading-order in system-bath coupling, as well as, respect all local conservation laws. In the above setting, it also captures thermalization of the isotropic XXZ-chain when all baths are at the same temperature. However, the RE can also be shown to not respect the CPTP property, and can lead to density matrix having negative eigenvalues of small magnitude. The question then is, whether there exists any physically consistent Markovian description, i.e., an LE which respects local conservation laws, gives correct populations and coherences to leading-order in system-bath coupling, and has CPTP property by definition. We use our SDP based formalism developed in Sec.~\ref{sec1-3} to answer this question.

 In particular, we get $\tau^{\rm pop}_{\rm opt}$ [Eq.~\eqref{eq:diag_optimization}] for understanding whether any LE can get correct populations of $ \rho^{(0)}_{\rm NESS}$, or get $\tau^{\rm pop,coh}_{\rm opt}$ [Eq.~\eqref{eq:coh_optimization}] for understanding whether the LE can get both correct populations and coherences of $ \rho^{(2)}_{\rm NESS}$. We construct the basis for operators in Hilbert space of the first $N_L$ and last $N_R$ qubits, so that the most general form of the desired LE can be written as in Eq.~\eqref{eq:local_lindblad_for_TOP}. For the $\ell$th qubit, we choose the orthonormal basis $\{ -\sigma^{(\ell)}_z /\sqrt{2}, \sigma^{(\ell)}_+, \sigma^{(\ell)}_-, I^{(\ell)}_2 / \sqrt{2} \} $, and $I^{(\ell)}_2$ is the identity operator for the qubit Hilbert space. The basis for the first $N_L$ and the last $N_R$ qubits is obtained by direct product of the basis of each of the qubits. The optimization problems are set in this basis, which we then directly input in the CVX MATLAB package to obtain $\tau^{\rm pop}_{\rm opt},~\tau^{\rm pop,coh}_{\rm opt}$. We set a value of the tolerance $\delta_{\rm tol}=10^{-6}$. \mkdev{The specific choice of $\delta_{tol} = 10^{-6}$ is directly dictated by the reliable precision limits of standard interior-point SDP solvers as the duality gap vanishes~\cite{BoydVandenberghe2004}.}


\subsection{Single qubit attached to left and right baths}
\label{sec:nl1}

First, we consider the case where the first and last qubit of the chain are coupled to the left and right baths, respectively, {which means } \( N_L = N_R = 1 \). We analyze the behavior of \( \tau^{\rm pop}_{\rm opt} \) and \( \tau^{\rm pop,coh}_{\rm opt} \) under different conditions using the isotropic XXZ Hamiltonian with \( N_M = 2 \), while keeping \( \beta_L = 1.0 \). The findings are summarized {below}.

Our analysis first considers the effect of varying the {inter-qubit} coupling strength $g$ for several values of $\beta_R$, as depicted in Figs.~\ref{fig:Nl1}(c,d) and \ref{fig:Nl1_app}(c,d). The results for the $N_L = N_R = 1$ case consistently show that $\tau^{\rm pop}_{\rm opt}$ and $\tau^{\rm pop, coh}_{\rm opt}$ are significantly greater than the tolerance $\delta_{\rm tol}$. This finding explicitly rules out the possibility of a general LE that can simultaneously preserve complete positivity, obey local conservation laws, and correctly capture the leading-order populations or coherences within this parameter range. We observe that both $\tau^{\rm pop}_{\rm opt}$ and $\tau^{\rm pop, coh}_{\rm opt}$ increase with $g$, which {generally} aligns with previous results~\cite{Tupkary_2022,archak} suggesting that local Lindblad equations provide a better description only when the system-qubit coupling is weak. {The accurate capture of these coherences is critical, as they play a significant role in quantum thermodynamic processes~\cite{Guarnieri_2018,Trushechkin_2022}.}

In contrast, the analysis of varying $\beta_L$ while keeping a fixed $g = 0.01$ reveals the same {behaviour of populations and coherences being above the tolerance $\delta_{\rm tol}$}, as shown in Figs.~\ref{fig:Nl1}(a,b) and \ref{fig:Nl1_app}(a,b). At both non-equilibrium and equilibrium, the values of $\tau^{\rm pop}_{\rm opt}$ and $\tau^{\rm pop, coh}_{\rm opt}$ are much larger than $\delta_{\rm tol}$. The conclusion is that constructing a fully consistent LE with only a single qubit coupled to baths at the left and right sites is not possible over a very large parameter regime. 


\begin{figure}
	\centering
	\includegraphics[width=\columnwidth]{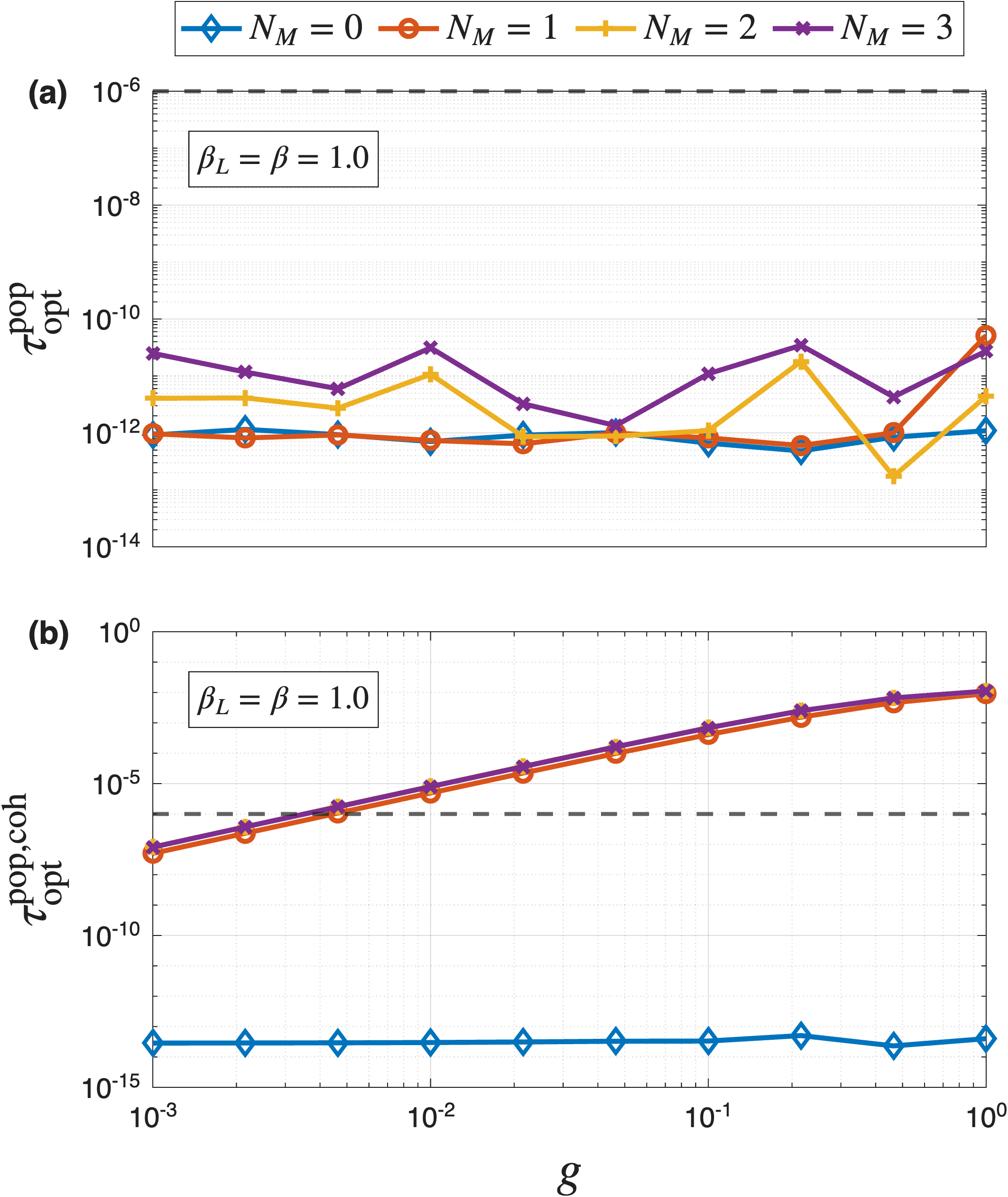}
\caption{Plots for (a) $\tau^{\rm pop}_{\rm opt}$ [Eq.~\eqref{eq:diag_optimization}]  and (b) $\tau^{\rm pop,coh}_{\rm opt}$ [Eq.~\eqref{eq:coh_optimization}] versus $g$ with $N_L = 3$, $N_R=0$,  $\beta_L = \beta = 1.0$ for isotropic XXZ Hamiltonian and four different values of $N_M =$ 0 (blue diamond), 1 (red circle), 2 (yellow cross), 3 (purple plus). The black dashed curve denotes $\delta_{\rm tol} = 10^{-6}$ with $\gamma_\ell = 1~\forall \ell,~\omega_c = 10$ [Eq.~\eqref{eq:ohmic_bath}].}
	\label{fig:cohg}
\end{figure}

\begin{figure}
	\centering
	\includegraphics[width=\columnwidth]{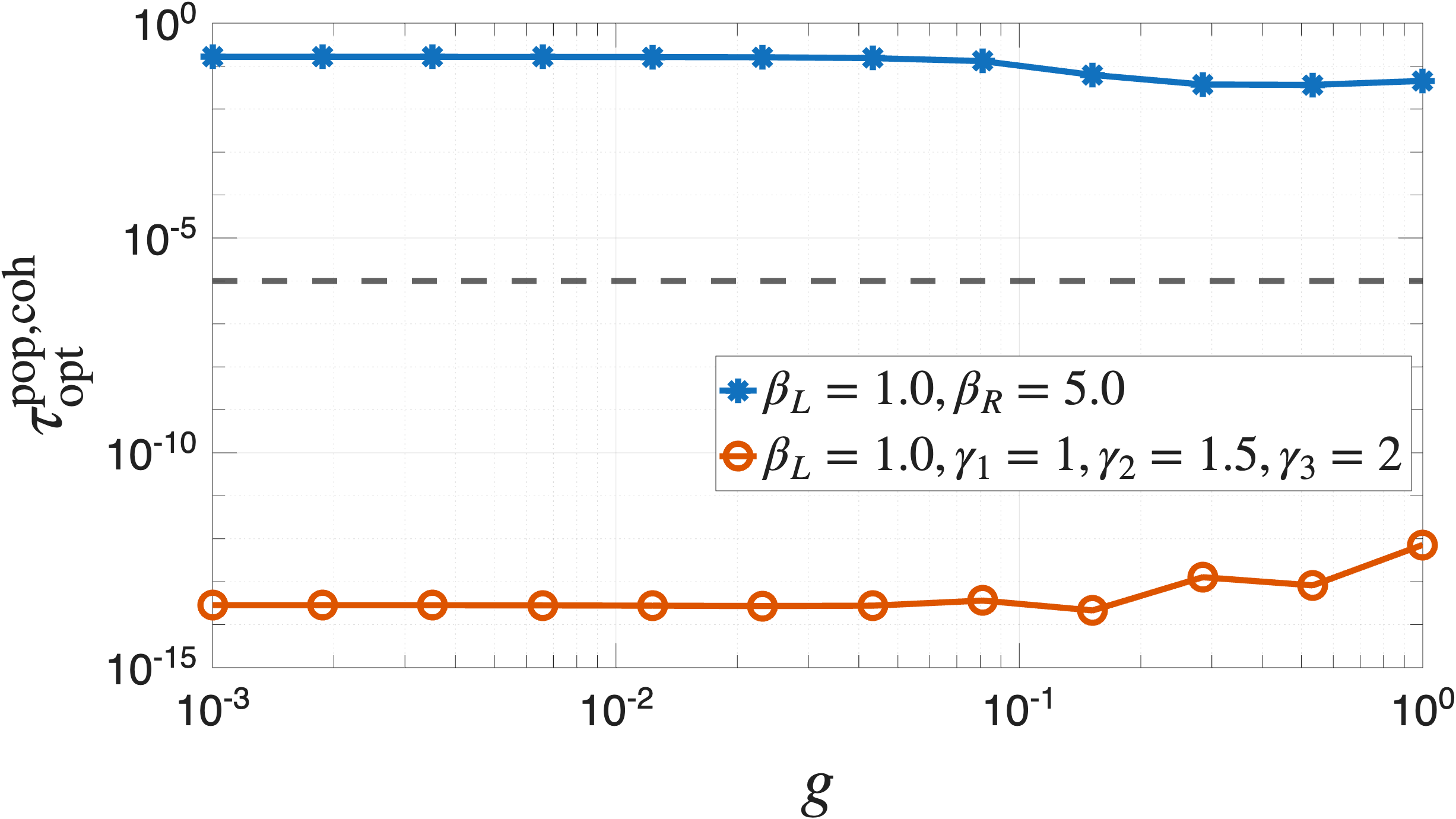}
	\caption{Plots for $\tau^{\rm pop,coh}_{\rm opt}$ [Eq.~\eqref{eq:coh_optimization}] with $N_L = 3,N_M = N_R = 0$ with $\beta_L = \beta = 1.0$, $\gamma_1 = 1,\gamma_2 = 1.5,\gamma_3=2$ (red circle) and $N_L = 1,N_M = 0,N_R = 2$ with $\beta_L = 1.0,~\beta_R = 5.0,~\gamma_\ell = 1~\forall \ell$ (blue star) versus $g$. The black dashed curve denotes $\delta_{\rm tol} = 10^{-6}$ and $\omega_c = 10$ [Eq.~\eqref{eq:ohmic_bath}].}
	\label{fig:cohg_comp}
\end{figure}

\subsection{Two qubits attached to left and right baths}
\label{sec:nl2}

We consider the case where the first and last two qubits of the chain are coupled to the left and right baths, respectively, {which means} \( N_L = N_R = 2 \). We analyze the behavior of \( \tau^{\rm pop}_{\rm opt} \) and \( \tau^{\rm pop,coh}_{\rm opt} \) under different conditions using the same parameters as in Sec.~\ref{sec:nl1}. \mkdev{Our primary motivation here is to investigate whether expanding the system-bath contact area can alleviate the strict no-go results observed for single-qubit contacts, as was observed in Ref.~\onlinecite{tupkary_searching_2023}.} We note that local two-qubit Lindblad dissipators have previously been employed to study energy transport in similar spin chains~\cite{znidaric_2011,Znidaric2015,Prosen_2009}. Our findings are summarized {below}.


{Our} results plotted in Figs.~\ref{fig:Nl2} and \ref{fig:Nl2_app}. We find a significant departure from the previous case in Sec.~\ref{sec:nl1} as $\tau^{\rm pop}_{\rm opt} \ll \delta_{\rm tol}$ for a relatively wider range of $g < 0.1$, as shown in Figs.~\ref{fig:Nl2}(c,d). This range increases as $\beta_L$ approaches $\beta_R$ (i.e. equilibrium). This suggests that, unlike the $N_L = N_R=1$ case, a LE that preserves complete positivity, obeys local conservation laws, and correctly captures zeroth-order populations may be possible even for non-equilibrium steady states. Despite the viability for populations, we find that $\tau^{\rm pop, coh}_{\rm opt} \gg \delta_{\rm tol}$
across most parameters, except at equilibrium or for very low $g$ values, indicating that coherences and populations \textit{together} remain difficult to capture.

Upon varying $\beta_L$ while keeping $g = 0.01$ (Figs.~\ref{fig:Nl2}(a,b) and \ref{fig:Nl2_app}(a,b)), non-monotonic trends in $\tau^{\rm pop}_{\rm opt}, \tau^{\rm pop,coh}_{\rm opt}$ {are} observed. We find $\tau^{\rm pop}_{\rm opt} \ll \delta_{\rm tol}$ across a wide range of $\beta_L$. It is observed that $\tau^{\rm pop, coh}_{\rm opt} \gg \delta_{\rm tol}$, except at equilibrium. This is consistent with previous observations, given the weak coupling strength. 

\subsection{Correct populations and coherences with baths attached to all qubits}

Here, we try a different setup from previous two subsections. First we keep $N_R = 0$ and attach Redfield baths of same temperature with the same coupling strength $\gamma_\ell = 1~\forall~l $ [Eq.~\eqref{eq:ohmic_bath}] to each of the first $N_L$ qubits. {In this setup, when $N_M = 0$, it is seen that all of our desired physical properties are satisfied for the equilibrium (all baths of same temperature) scenario.}

Fig.~\ref{fig:cohg} highlights the results of this optimization by plotting $\tau^{\rm pop}_{\rm opt}$ {(upper panel)} and $\tau^{\rm pop,coh}_{\rm opt}$ {(lower panel)} versus different values of $g$. Most interestingly, we notice that when all the qubits of the system are connected to baths, it might be possible for the CVX-optimized Linbladian baths to get the correct {populations and} coherences for its steady state, as $\tau^{\rm pop,coh}_{\rm opt} \ll \delta_{\rm tol}$. We remark that we get similar results when we plot $\tau^{\rm pop,coh}_{\rm opt},\tau^{\rm pop}_{\rm opt}$ versus $\beta_L = \beta$ for {fixed} $g = 0.01$ and the other parameters being the same.

For the case of $N_M = N_R = 0$ in Fig.~\ref{fig:cohg}, one might consider the equilibrium setup with all coupling strengths same to be a very special scenario, and wish to see how robust the equilibrium case is when the bath coupling strengths are different across the qubits. One might also question what happens in the non-equilibrium case of all sites attached to baths (i.e. when $N_M = 0,N_R \neq 0$ and $\beta_L \neq \beta_R$). Fig.~\ref{fig:cohg_comp} highlights these plots, showing that while the equilibrium case still gets all the desired physical properties even with different coupling strengths of the baths to the qubits, in the non-equilibrium scenario we again are unable to satisfy correct populations and coherences together.

\section{Conclusions and Outlook}
\label{sec:discussion}

The results presented in this work demonstrate the power of convex optimization, specifically semidefinite programming, in probing the physical consistency of quantum master equations. Our analysis reveals a range of parameter regimes, particularly in the presence of strong {asymmetry in bath temperatures ($\beta_L \neq \beta_R$)} or large inter-site couplings where no LE can simultaneously satisfy complete positivity, preserve local conservation laws, and yield the correct NESS populations and coherences ({up to leading-order, see Table~\ref{Table:figs_main}}). While we focused on these {quantities in this study}, we expect that other properties may be investigated in a similar manner, as long as they can be expressed as a suitable optimization problem. Moreover, albeit we focused on the isotropic XXZ qubit chain, our approach is readily adaptable to other Hamiltonian setups. \mkdev{We emphasize that this work does not imply that Markovian QMEs lacking leading-order accuracy are without practical value, but simply highlights a paradigm to investigate whether a Markovian description can satisfy all fundamental requirements, and simultaneously checks it for our chosen few-qubit setups.}

One of the central insights from our approach is the identification of the trade-offs between {obtaining correct populations and} coherences ({up to leading-order}), and the enforcement of conservation laws in Markovian models. While equilibrium configurations with low coupling strength showed marginal feasibility, {the} non-equilibrium setups failed to admit consistent Lindblad descriptions, validating earlier theoretical predictions~\cite{tupkary_searching_2023}. 

\mkdev{A primary limitation of our exact numerical approach is its exponential computational scaling due to $\mathcal{O}(2^N)$ global constraints and basis transformations, rendering it unfeasible for larger systems ($N > 6$). However, to establish rigorous bounds on the metrics, one only needs to evaluate feasible points for the primal or dual problems, rather than completing a full global optimization. Because evaluating feasibility relies on local expectation values and superoperator applications, it maybe amenable to approximate many-body methods like Tensor Network ansatzes (such as Matrix Product Operators~\cite{prosen_2011,process_tensor}). This is an interesting avenue for future work.}

Looking forward, this semidefinite programming (SDP) - based approach can be extended in various directions by formulating other desired properties, or combinations thereof, within the framework of SDPs. Importantly, such an SDP formulation provides not only candidate optimal solutions but also reliable and tight optimal values, via the associated dual problem. This yields rigorous and quantitatively reliable bounds on how closely the desired properties can be satisfied. An especially important regime is when the solution of the SDP certifies that the desired properties cannot be simultaneously satisfied, even approximately, thereby yielding rigorous no-go statements rather than merely constructive approximations. Moreover, these guarantees can be accessed in practice without explicitly constructing dual problems or establishing duality results by hand: modern SDP solvers implement primal–dual algorithms internally and automatically return optimal values. As a result, much of the analytical burden can be outsourced to well-developed numerical tools, allowing the focus to remain on formulating physically meaningful metrics and constraints. 


Moreover, the emergence of quantum hardware for {Noisy Intermediate Scale Quantum} (NISQ) simulations offers an exciting avenue to validate these predictions experimentally, particularly in platforms capable of engineered dissipation such as trapped ions~\cite{Barreiro_2011}, Rydberg atoms~\cite{Schindler_2013}, superconducting circuits~\cite{Garcia_2020}, and semiconductor quantum dots~\cite{Kim_2022}, by comparing engineered Lindbladian dynamics to Redfield or microscopic evolution. Thus, our results present an interesting application of SDPs towards understanding open quantum system dynamics and lay the foundation for future explorations into controlled dissipative quantum engineering.


All code used in this work can be found in Ref.~\onlinecite{lo568los}.

\section{Acknowledgements}
We thank A. Dhar for useful discussions about this project. SS thanks the International Centre for Theoretical Sciences for its hospitality and acknowledges support via the Long Term Visiting Student Program (LTVSP). MK acknowledges the support of the Department of Atomic Energy, Government of India, under Project No. RTI4001.  MK thanks the hospitality of the Department of Physics, Princeton University. Part of this work
was performed at the Institute for Quantum Computing,
at the University of Waterloo, which is supported by Innovation, Science, and Economic Development Canada.
DT is supported
by the Mike and Ophelia Lazaridis Fellowship. AP thanks IIT Hyderabad Seed Grant for support.\\ 

\appendix

\setcounter{figure}{0}
\renewcommand{\thefigure}{A\arabic{figure}}

\section{Calculation of $\rho_{\rm NESS}^{(0)}$ from Redfield Equation}
\label{appn1}

In this appendix, we {derive the precise form of} 
\begin{align} \label{red_ness}
	\rho_{\rm NESS}^{(0)} = \sum_{a=1}^d p_a \ket{E_a} \bra{E_a}.
\end{align}
{Here $d = 2^N$ is the dimension of our total system}. We use the vector $\mathbf p = (p_1, p_2, \dots, p_\alpha \dots)^T$ and note that the second-order dissipator (${\mathcal{L}}_2$) of the RE is \cite{Tupkary_2022}
\begin{align} \label{l2_red}
	{\mathcal{L}}_2(\rho) &= \sum_\ell  \sum_{\alpha, \gamma = 1}^{2^N}\{[\rho \proj{E_\alpha} \sigma_-^{(\ell)} \proj{E_\gamma},\sigma_+^{(\ell)}]C_\ell(\alpha,\gamma) \notag\\
	&+ [\sigma_+^{(\ell)}, \proj{E_\alpha}\sigma_-^{(\ell)}\proj{E_\gamma}]D_\ell(\alpha,\gamma) + \text{H.c.}\},
\end{align}
with
\begin{equation}
	\begin{aligned}
		C_{\ell}(\alpha,\gamma) &= \frac{\mathfrak{J}_{\ell}(E_{\gamma \alpha}) n_{\ell}(E_{\gamma \alpha})}{2 } - i \mathcal{P} \int_{0}^{\infty} d \omega \frac{\mathfrak{J}_{\ell}(\omega) n_{\ell}(\omega)}{\omega-E_{\gamma \alpha}},  \\
		D_{\ell}(\alpha,\gamma) &= \frac{ e^{\beta_{\ell}(E_{\gamma \alpha}-\mu_{\ell})} \mathfrak{J}_{\ell}(E_{\gamma \alpha}) n_{\ell}(E_{\gamma \alpha})}{2 }\\ &- i \mathcal{P} \int_{0}^{\infty} d \omega \frac{e^{\beta_{\ell}(\omega-\mu_{\ell})} \mathfrak{J}_{\ell}(\omega) n_{\ell}(\omega)}{\omega-E_{\gamma \alpha}},
		\label{redfield:constants}
	\end{aligned}
\end{equation}
where $\mathcal{P}$ denotes the Cauchy Principal value, {$E_{\gamma \alpha} = E_\gamma - E_\alpha$, $n_\ell(\omega) = (e^{\beta_\ell(\omega - \mu_\ell)}-1)^{-1}$ is the average occupation number and $\mathfrak{J}_\ell(\omega)$ is the spectral bath function. We take an ohmic form $\mathfrak{J}_\ell(\omega) = \omega \exp(-(\omega/\omega_c)^2)$ for the spectral bath function}. {The summation over} $\ell$ {in Eq.~\eqref{l2_red} is performed over} the qubits where the baths are connected ($\ell = 1,N$ for $N_L = N_R = 1$ and $\ell = 1,2,N-1,N$ for $N_L = N_R = 2$) and $H_S \ket{E_\alpha} = E_\alpha \ket{E_\alpha}$. Now, from Eq.~\eqref{diagonal_values} {of the main text}, we know that $\bra{E_k} {\mathcal{L}}_2[\rho_{\rm NESS}^{(0)}] \ket{E_k} = 0~~\forall~k$. We {thus} have
\begin{widetext}
	\begin{align} \label{lineq_re}
		\bra{E_k} \rho_{\rm NESS}^{(0)} \ket{E_k} &=  \sum_{\alpha, \gamma = 1}^{2^N} \sum_\ell \{p_\alpha \bra{E_k}[\proj{E_\alpha} \sigma_-^{(\ell)} \proj{E_\gamma}, \sigma_+^{(\ell)}]\ket{E_k} C_\ell(\alpha,\gamma) \notag\\ 
		&+ p_\gamma \bra{E_k}[\sigma_+^{(\ell)}, \proj{E_\alpha}\sigma_-^{(\ell)}\proj{E_\gamma}]\ket{E_k} D_\ell(\alpha,\gamma) + \text{H.c.}\}.
	\end{align} 
\end{widetext}

In Eq.~\eqref{lineq_re}, the L.H.S is 0, and the R.H.S can be written as a vector dot product. {More precisely,} we can rewrite Eq.~\eqref{lineq_re} as {some} matrix $A$ multiplied with vector $\mathbf{p}$ giving the $\mathbf{0}$ vector {($A \mathbf p = \mathbf 0$)}. {The matrix elements $A_{km}$ is given } by {collecting} the coefficients of $p_m$ in the R.H.S of Eq.~\eqref{lineq_re}, giving us
\begin{widetext}
	\begin{align}
		A_{km} &= \sum_{\gamma = 1}^{2^N} \sum_{\ell} \big\{ (\bra{E_k} [\proj{E_m} \sigma_-^{(\ell)} \proj{E_\gamma},\sigma_+^{(\ell)}]\ket{E_k} C_\ell(m,\gamma) + \text{H.c.}) \notag \\
		&+ (\bra{E_k}[\sigma_+^{(\ell)}, \proj{E_\gamma}\sigma_-^{(\ell)}\proj{E_m}]\ket{E_k} D_\ell(\gamma,m) + \text{H.c.}) \big\}.
	\end{align}
\end{widetext}
Along with the constraints arising from $p_a$ being the elements of a diagonal density matrix ($\sum_a p_a = 1,~0 \leq p_a \leq 1$), we can solve {the equation $A \mathbf p = \mathbf 0$} using standard linear equation solving techniques and arrive at $\vect{p}$. {This} gives us the required diagonal elements of $\rho^{(0)}_{\rm NESS}$, presented in Eq.~\eqref{red_ness}.

\begin{table*}[t]
	\centering
	\begin{tabular}{|c|c|c|c|c|c|c|c|c|}
		\hline
		$N_L = N_R$  & $\tau^{\rm pop}_{\rm opt}$ v/s $\beta$ &$\tau^{\rm pop,coh}_{\rm opt}$ v/s $\beta$ & $\tau^{\rm pop}_{\rm opt}$ v/s $g$ & $\tau^{\rm pop,coh}_{\rm opt}$ v/s $g$ & Correct populations & Correct coherences & Local conservation laws\\ \hline \hline
		
		1   & Fig.~\ref{fig:Nl1_app}(a) & Fig.~\ref{fig:Nl1_app}(b)& Fig.~\ref{fig:Nl1_app}(c) & Fig.~\ref{fig:Nl1_app}(d) & Impossible & Impossible & Satisfied\\ \hline
		
		2   & Fig.~\ref{fig:Nl2_app}(a) & Fig.~\ref{fig:Nl2_app}(b)& Fig.~\ref{fig:Nl2_app}(c)& Fig.~\ref{fig:Nl2_app}(d) & Possible & Impossible & Satisfied\\ \hline
		
	\end{tabular}
	\caption{\label{Table:figs} Table of figure references {summarising our results in the non-equilibrium ($\beta_L \neq \beta_R$) scenario} for different energy biases and number of qubits $N_L,~N_R$ attached to the left and right baths {respectively} in the biased case of $\epsilon_0 = 0.01$ [Sec.~\ref{appn3}]. All plots are done for system-bath coupling $\epsilon = 0.01$.}
\end{table*}

\begin{figure*}
	\centering
	\hspace*{-10mm}
	\includegraphics[width=2\columnwidth]{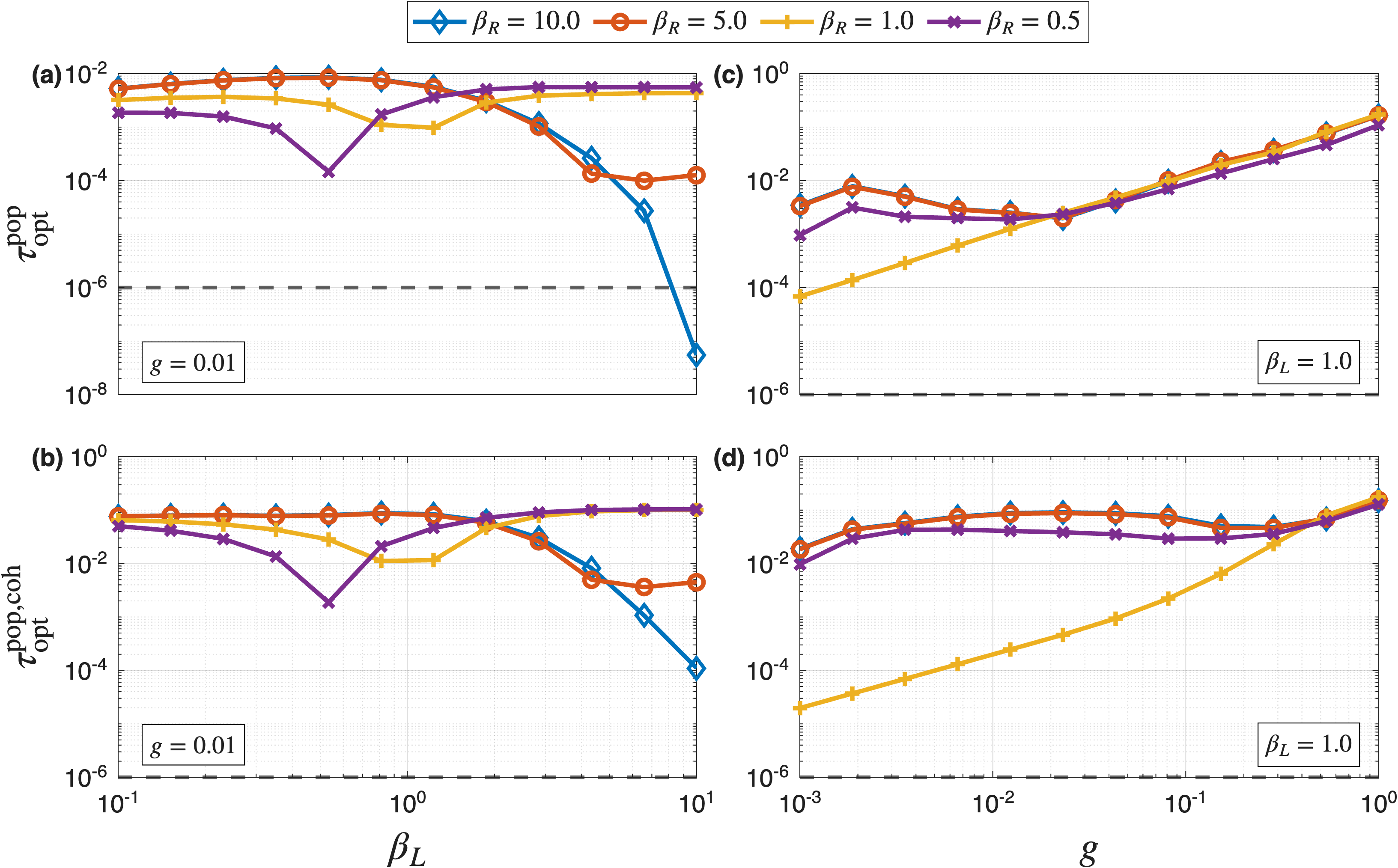}
	\caption{Plots for $N_L = N_R = 1,~N_M = 2$ (i.e. single qubit attached to left and right baths, see Sec.~\ref{sec:nl1}) for the isotropic XXZ Hamiltonian [Eq.~\eqref{eq:ham}] keeping $\gamma_\ell = 1 ~\forall \ell,~\omega_c=10$ [Eq.~\eqref{eq:ohmic_bath}], $\epsilon_0 = 0.01$~[App.~\ref{appn3}], and $\beta_R = 10.0$ (blue diamond), $5.0$ (red circle), $1.0$ (yellow cross), $0.5$ (purple plus). {The black dashed horizontal line represents $\delta_{\rm tol} = 10^{-6}$}.(a) $\tau^{\rm pop}_{\rm opt}$ [Eq.~\eqref{eq:diag_optimization}] versus $\beta_L$ with $g=0.01$. (b) $\tau^{\rm pop,coh}_{\rm opt}$ [Eq.~\eqref{eq:coh_optimization}] versus $\beta_L$ with $g=0.01$. (c) $\tau^{\rm pop}_{\rm opt}$ versus $g$ for $\beta_L = 1.0$. (d) $\tau^{\rm pop,coh}_{\rm opt}$ versus $g$ for $\beta_L = 1.0$. This figure shows that we can never obtain even the correct leading-order populations (implying incorrect leading-order coherences as well) with only a single qubit attached at left and right baths.}
	\label{fig:Nl1_app}
\end{figure*}

\begin{figure*}
	\centering
	\hspace*{-10mm}
	\includegraphics[width=2\columnwidth]{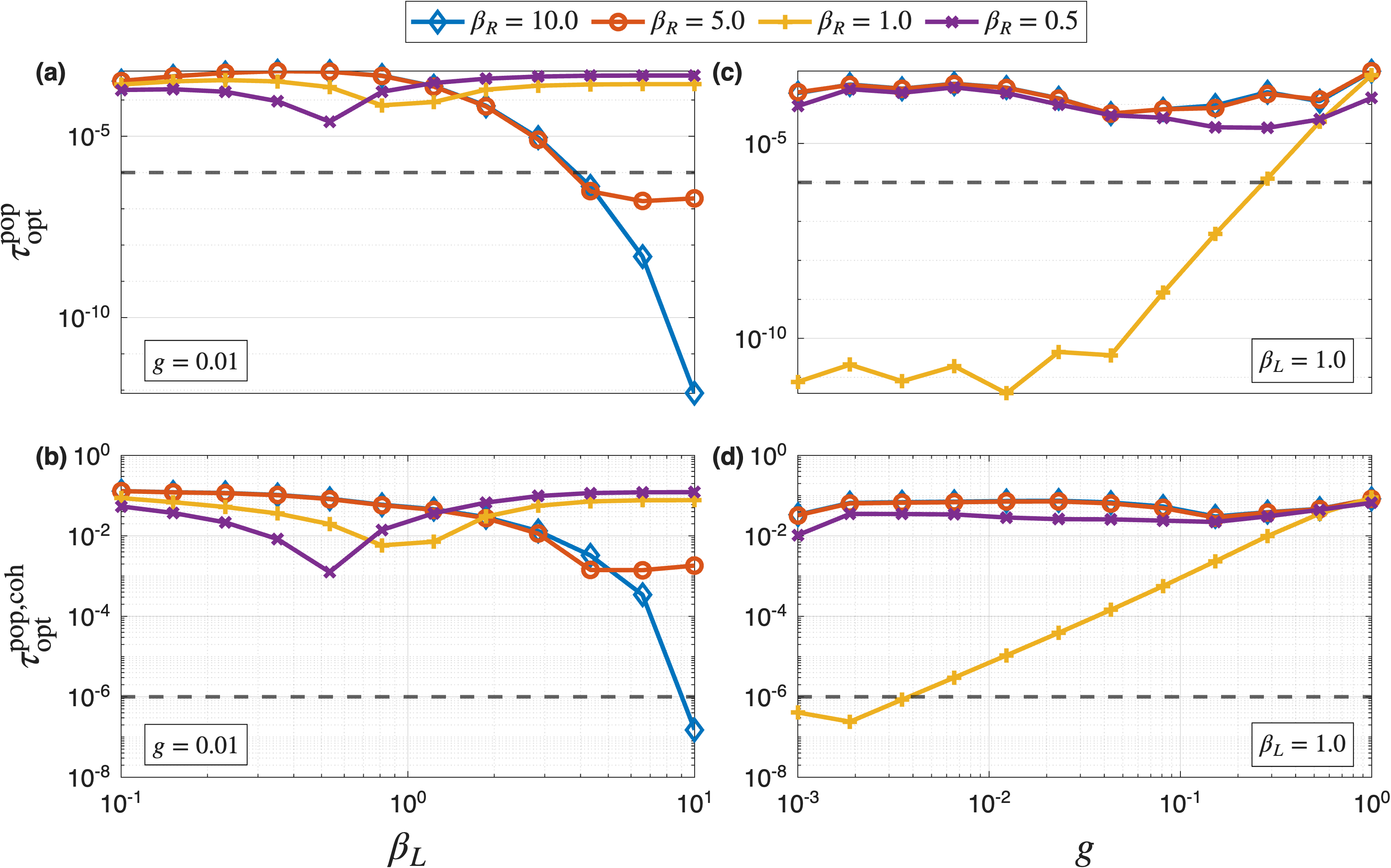}
	\caption{Plots for $N_L = N_R = 2,~N_M = 2$ for the isotropic XXZ Hamiltonian [Eq.~\eqref{eq:ham}] keeping $\gamma_\ell = 1 ~\forall \ell,~\omega_c=10$ [Eq.~\eqref{eq:ohmic_bath}], $\epsilon_0 = 0.01$~[App.~\ref{appn3}], and $\beta_R = 10.0$ (blue diamond), $5.0$ (red circle), $1.0$ (yellow cross), $0.5$ (purple plus). (a) $\tau^{\rm pop}_{\rm opt}$ [Eq.~\eqref{eq:diag_optimization}] versus $\beta_L$ with $g=0.01$. (b) $\tau^{\rm pop,coh}_{\rm opt}$ [Eq.~\eqref{eq:coh_optimization}] versus $\beta_L$ with $g=0.01$. (c) $\tau^{\rm pop}_{\rm opt}$ versus $g$ for $\beta_L = 1.0$. (d) $\tau^{\rm pop,coh}_{\rm opt}$ versus $g$ for $\beta_L = 1.0$. This figure shows that for a system with a small bias for onsite energies, even correct leading-order populations are unattainable in a wide parameter regime (implying incorrect leading-order coherences in these regimes as well).}
	\label{fig:Nl2_app}
\end{figure*}

\mkdev{\section{Accurate leading-order coherences implies leading-order local conservation.}
\label{appn4}
In the appendix, we prove the claim that for a generic QME, accurate leading-order coherences in the NESS implies local conservation laws. We take the generic QME of the form
\begin{align}
	\label{TCL2New_app}
	\frac{\partial \cvxrho}{\partial t}= {\mathcal{L}}_{0}[\cvxrho(t)]+ \epsilon^2 {\mathcal{L}}^\prime_{2}[\cvxrho(t)].
\end{align}
}\mkdev{We take a time-independent operator $O$ living in the system's Hilbert space $\mathcal H_S$, commuting with the system-bath operator $H_{SB}$ [Eq.~\eqref{eq:system_ham}], and given that the generic NESS $\cvxrho_{\rm NESS}$ has accurate leading-order (i.e. up to $\epsilon^2$) coherences, which means via Eq.~\eqref{offdiagonal_condition_og}
\begin{align} \label{eq:offdiagonal_og_appn}
 \bra{E_a}\mathcal L'_2[\cvxrho^{(0)}_{\rm NESS}] \ket{E_\nu} = \bra{E_a}\mathcal L_2[\genrho^{(0)}_{\rm NESS}] \ket{E_\nu}~~\forall E_a \neq E_\nu,   
\end{align}
where $H_S = \sum_a E_a \ket{E_a}\bra{E_a},~\mathcal L_2$ is the RE second-order Liouvillian.} 

\mkdev{To prove that we have local conservation in this case, we go to the necessary and sufficient condition for satisfying local conservation laws for any QME of the form Eq.~\eqref{TCL2New_app}
\begin{align}
    \Tr[O\mathcal L'_2[\cvxrho(t)]] = 0,
\end{align}
or for order-by-order for $\cvxrho_{\rm NESS} = \sum_m\epsilon^{2m}\cvxrho^{(m)}_{\rm NESS}$
\begin{align}
    \Tr[O\mathcal L'_2[\cvxrho^{(m)}_{\rm NESS}]] = 0.
\end{align}
This condition is already known to be true in the case of RE~\cite{Tupkary_2022,tupkary_searching_2023} up to all orders, that is 
\begin{align} \label{eq:re_loccons}
    \Tr[O\mathcal L_2[\genrho(t)]] = 0 \implies \Tr[O\mathcal L_2[\genrho^{(0)}_{\rm NESS}]] = 0.
\end{align}
Expanding out the operator evolution for the generic QME state $\cvxrho_{\rm NESS}$ up to $\epsilon^2$ gives us}
\mkdev{\begin{widetext}
\begin{align}
 \frac{d \langle O \rangle_{\cvxrho_{\rm NESS}}}{dt} = \Tr[O\frac{\partial \cvxrho_{\rm NESS}}{\partial t}] &= -i\langle [O,H_S]\rangle_{\cvxrho_{\rm NESS}} + \epsilon^2\Tr[O\mathcal L'_2[\cvxrho^{(0)}_{\rm NESS}]].   
\end{align}
Expanding the second trace term out in the energy eigenbasis gives us
\begin{align}
 &\Tr[O\mathcal L'_2[\cvxrho^{(0)}_{\rm NESS}]] =   \sum_{a,\nu=1}^D \bra{E_\nu}O\ket{E_a}\bra{E_a} L'_2[\cvxrho^{(0)}_{\rm NESS}]\ket{E_\nu} \notag \\ \label{eq:sum_decomp}
 &=\sum_{a \neq \nu}^D \bra{E_\nu}O\ket{E_a}\bra{E_a} L'_2[\cvxrho^{(0)}_{\rm NESS}]\ket{E_\nu} + \sum_{a =1}^D \bra{E_a}O\ket{E_a}\bra{E_a} L'_2[\cvxrho^{(0)}_{\rm NESS}]\ket{E_a}.
\end{align}
\end{widetext}
In the first sum of Eq.~\eqref{eq:sum_decomp}, we can replace $\bra{E_a} L'_2[\cvxrho^{(0)}_{\rm NESS}]\ket{E_\nu}$ with $\bra{E_a} L_2[\genrho^{(0)}_{\rm NESS}]\ket{E_\nu}$ [Eq.~\eqref{eq:offdiagonal_og_appn}]. In the second sum of Eq.~\eqref{eq:sum_decomp}, we can also replace $\bra{E_a} L'_2[\cvxrho^{(0)}_{\rm NESS}]\ket{E_a}$ with $\bra{E_a} L_2[\genrho^{(0)}_{\rm NESS}]\ket{E_a}$ (as both are equivalently zero, see Eq.~\eqref{diagonal_values}
in the main text). Then we have that 
\begin{align}
  \Tr[O\mathcal L'_2[\cvxrho^{(0)}_{\rm NESS}]] =   \Tr[O\mathcal L_2[\genrho^{(0)}_{\rm NESS}]] = 0,
\end{align}
where the last equality comes from Eq.~\eqref{eq:re_loccons}, which gives us the leading-order sufficient condition for local conservation. This implies that, up to leading-order, accurate coherences implies local conservation in a generic QME.}

\section{Trace distance between the obtained and the zeroth-order steady state}
\label{appn2}

In this section, we prove \cref{lm:trace_dist} of the main text. First, we recall Eq.~\eqref{eq:diag_optimization}
\begin{equation} \label{tauequation_app}
	\tau^{\rm pop} = \sum_\alpha \lvert  \bra{ E_{\alpha} } {\mathcal{L}}^\prime_2[\genrho_{\rm NESS}^{(0)}] \ket{ E_{\alpha} } \rvert,
\end{equation}
and recall Lemma~\ref{lm:trace_dist}.
\tracedistlem*

\begin{proof}
	Recall from Eq.~\eqref{red_ness} that $\rho_{\rm NESS}^{(0)} = \sum_a p_a \ket{E_a} \bra{E_a}$ is the non-equilibrium steady state obtained from RE with second-order dissipator ${\mathcal{L}}_2$ as per Appendix~\ref{appn1} [Eq.~\eqref{l2_red}]. Here, recall that $\{\ket{E_a}\}$ is the energy eigenbasis of our system Hamiltonian $H_S$. On the other hand, $\cvxrho_{\rm NESS}^{(0)} = \sum_a p^\prime_a \ket{E_a} \bra{E_a}$ is the zeroth-order steady state of the master equation with the general Lindbladian second-order dissipator ${\mathcal{L}}_2^\prime$ as given in Eq.~\eqref{eq:local_lindblad_for_TOP}. We have from Eq.~\eqref{diagonal_values}, 
	\begin{align}
		\bra{E_k} {\mathcal{L}}_2(\rho_{\rm NESS}^{(0)}) \ket{E_k} = 0~~\forall~k 
	\end{align}
	Also, from the solution to the optimization problem, we know that
	\begin{align} 
		&\sum_k|\bra{E_k} {\mathcal{L}}^\prime_2(\rho_{\rm NESS}^{(0)}) \ket{E_k}| \geq \tau^{\rm pop}_{\rm opt},
	\end{align}
	which implies, after substituting $\rho_{\rm NESS}^{(0)}$ using Eq.~\eqref{red_ness},
	\begin{align}
		\label{app2_eq2}
		&\sum_k \bigg| \sum_a p_a \bra{E_k} {\mathcal{L}}^\prime_2(\proj{E_a}) \ket{E_k}\bigg| \geq \tau^{\rm pop}_{\rm opt}.
	\end{align}
	Let us define a matrix $C$ such that its elements are given by
	\begin{align}
		C_{ka} \coloneqq  \bra{E_k} {\mathcal{L}}^\prime_2(\proj{E_a}) \ket{E_k}.
	\end{align}
	We can then rewrite Eq.~\eqref{app2_eq2}, using $b_k \coloneq \sum_a  C_{ka}p_a$, as
	\begin{align} \label{b7}
		\norm{C\vect{p}}_1 = \sum_k \bigg| \sum_a  C_{ka}p_a \bigg| =\sum_k|b_k| \geq \tau^{\rm pop}_{\rm opt} , 
	\end{align}
	where the subscript 1 in the L.H.S. denotes the $\text{L}_1~\text{norm}$. Eq.~\eqref{b7} can by compactly written as 
	\begin{align}
		\norm{C\vect{p}}_1 = \norm{\vect{b}}_1,\quad \vect{p} = (\{p_a\})^T,
	\end{align} 
	where $\vect{b} = (\{b_k\})^T$. In the same manner taking $\vect{p^\prime} = (\{p^\prime_a\})^T$, we see that
	\begin{align}
		{\mathcal{L}}^\prime_2(\cvxrho_{\rm NESS}^{(0)}) = 0 \implies C\vect{p^\prime} = 0. 
	\end{align}
	Our central question can now be reformulated as finding the upper and lower bounds on the quantity $\norm{\vect{p}-\vect{p^\prime}}_1$, given that $\norm{C\vect{p}}_1 \geq \tau^{\rm pop}_{\rm opt},~C\vect{p^\prime}=0$, with $\vect{p,~p'}$ being probability vectors. It is important to emphasize that this expression ($\norm{\vect{p}-\vect{p^\prime}}_1$) is identical to the trace distance between the two diagonal density matrices $\rho_{\rm NESS}^{(0)} = \sum_a p_a \ket{E_a} \bra{E_a}$ and $\cvxrho_{\rm NESS}^{(0)} = \sum_a p^\prime_a \ket{E_a} \bra{E_a}$. 
	
	We now utilize the triangle inequality, to show that for any vector $\vect{l}$, we have
	\begin{equation} \label{b10}
		\begin{aligned}
			\norm{C\vect{l}}_1 &=
			\sum_k \bigg| \sum_a C_{ka}l_a \bigg|  \\
			&\leq \sum_a  \sum_k  \left( |l_a C_{ka}| \right) \\
			&= \sum_a \left( |l_a| \sum_k |C_{ka}| \right).
		\end{aligned}
	\end{equation}
	The inner sum $\sum_k |C_{ka}|$ in the RHS of Eq.~\eqref{b10} corresponds to the total ``weight" applied by the matrix $C$ to each component $|l_a|$. Let us define a constant $\alpha$  which bounds $\sum_k |C_{ka}|$ as %
	\begin{align} \label{b11}
		\sum_k |C_{ka}\big| \leq \alpha. 
	\end{align}
	Now, substituting $\vect{l} = \vect{p} - \vect{p^\prime}$ and noting that $C\vect{p^\prime} = 0,~ \norm{C\vect{p}}_1 \geq \tau^{\rm pop}_{\rm opt}$, we have from Eqs.~\eqref{b10} and \eqref{b11}
	\begin{equation}
		\begin{aligned}
			\norm{\vect{p} - \vect{p^\prime}}_1 &\geq \norm{C(\vect{p} - \vect{p^\prime})} / \alpha \\
			&= \norm{C\vect{p}} / \alpha  \\
			&\geq
			\tau^{\rm pop}_{\rm opt}/\alpha.    
		\end{aligned}
	\end{equation}
	The only thing left to do in our proof is to find an  explicit form of the upper bound $\alpha$. We will do so with some straightforward algebra. Given that ${\mathcal{L}}_2^\prime$ contains $\Gamma^{(L)}$ and $\Gamma^{(R)}$ [Eq.~\eqref{eq:local_lindblad_for_TOP}], we have
	\begin{widetext}
		\begin{align} \label{b13}
			C_{ka} &= \bra{E_k} {\mathcal{L}}_2^\prime(\proj{E_a}) \ket{E_k} \notag \\
			&= \sum_{ij}^{d_L^2 - 1} \frac{\Gamma^{(L)}_{ij}}{2^{(N_M + N_R)}}(\bra{E_k}(f_i \otimes I_{MR}) \proj{E_a}(f_j \otimes I_{MR})^\dag \ket{E_k} - \bra{E_k}(f_i \otimes I_{MR})^\dag(f_j \otimes I_{MR})\ket{E_k} \delta_{ka}) \notag \\
			&+\sum_{ij}^{d_R^2 - 1} \frac{\Gamma^{(R)}_{ij}}{2^{(N_M + N_R)}}(\bra{E_k}(I_{LM} \otimes f_i) \proj{E_a}(I_{LM} \otimes f_j)^\dag \ket{E_k} - \bra{E_k}(I_{LM} \otimes f_i)^\dag(I_{LM} \otimes f_j)\ket{E_k} \delta_{ka})
		\end{align} 
	\end{widetext}
	The $H_{LS}$ terms in Eq.~\eqref{eq:local_lindblad_for_TOP} do not contribute as they are Hermitian and hence cancel off after expanding the commutator. To analyze Eq.~\eqref{b13} further, we concentrate only on the term with the left bath sum (second line in Eq.~\eqref{b13}). A similar following analysis also holds for the right bath sum (third line in Eq.~\eqref{b13}). We define
	\begin{align}
		m_{ka}^i = \bra{E_k} (f_i \otimes I_{MR}) \ket{E_a}.  
	\end{align}
	The elements $m_{ka}^i$ are just the value at the $(k,a)$ position of the matrix $m^i\coloneqq f_i \otimes I_{MR}$ in the energy eigenbasis. The idea now is to find the operator norm of $m^i$ (denoted by $\norm{m^i}$) and use the fact that the absolute value of a matrix element in an orthonormal basis is bounded by the operator norm of the matrix~\cite{horn2012matrix}. In other words, we use the fact that 
	\begin{align} \label{eq:tempmi}
		|m^i_{ka}| \leq \norm{m^i}.
	\end{align}
	The operator norm of a matrix is simply the largest absolute value of the singular values of matrix, and the matrices in $m^i$ are tensor products of simpler, smaller matrices. In particular, we have $\norm{-\sigma_z/\sqrt{2}} = \norm{I_2/\sqrt{2}} = 1/\sqrt 2$, and  $\norm{\sigma_+} = \norm{\sigma_-} = 1$. Now, $f_i$ is a tensor product over different basis operators, that is  $f_i \in \{-\sigma_z/\sqrt{2},\sigma_+,\sigma_-,I_2/\sqrt{2}\}^{\otimes N_L}$. Using the fact that $\norm{A \otimes B} = \norm{A} \norm{B}$, and Eq.~\eqref{eq:tempmi}, we obtain 
	
	\begin{align} \label{eq:mibound}
		|m_{ka}^i| \leq \norm{m^i} \leq 1.
	\end{align}
	A similar analysis holds for the right bath, which we denote using $m'$ instead of $m$.
	Iteratively applying triangle inequalities on Eq.~\eqref{b13}, alongside our modified notation, gives us
	\begin{widetext}
		\begin{align} \label{b17}
			\sum_k|C_{ka}| &= \sum_k\bigg|\sum_{ij} \frac{\Gamma^{(L)}_{ij}}{2^{(N_M + N_R)}}\left(m^i_{ka}(m^j_{ka})^\dag - \sum_n(m^i_{nk})^\dag m^j_{nk}\delta_{ka}\right) \notag + \frac{\Gamma^{(R)}_{ij}}{2^{(N_M + N_L)}}\left(m'^i_{ka}(m'^j_{ka})^\dag - \sum_n(m'^i_{nk})^\dag m'^j_{nk}\delta_{ka}\right)\bigg| \notag \\
			&\leq \sum_k \sum_{ij} \frac{\big| \Gamma^{(L)}_{ij} \big| }{2^{(N_M + N_R)}}\left( |m^i_{ka}(m^j_{ka})^\dag | + \sum_n | (m^i_{nk})^\dag m^j_{nk}| \delta_{ka}\right) \notag + \frac{\big| \Gamma^{(R)}_{ij}\big| }{2^{(N_M + N_L)}}\left( | m'^i_{ka}(m'^j_{ka})^\dag |  + \sum_n | (m'^i_{nk})^\dag m'^j_{nk} | \delta_{ka}\right)\bigg| \notag \\   
			& \leq \left(\sum_{ij}\big| \Gamma^{(L)}_{ij}\big| \left(\frac{d_L}{d}\right){\sum_k  (1 + d\delta_{ka})} + \sum_{ij}\big| \Gamma^{(R)}_{ij}\big|\left(\frac{d_R}{d}\right){\sum_k (1 + d\delta_{ka})}\right) \notag \\
			&= 2\left( \sum_{ij}\big| \Gamma^{(L)}_{ij}\big|d_L + \sum_{ij}\big| \Gamma^{(R)}_{ij}\big|d_R\right)   
		\end{align}   
	\end{widetext}
	%
	%
	%
	where $d_L = 2^{N_L},~d_R = 2^{N_R},~d_M = 2^{N_M},~d = d_Ld_Md_R$. Here, we used Eq.~\eqref{eq:mibound} in the third line. For the fourth line, we use the fact that $\Gamma^{(L)},~\Gamma^{(R)}$ are positive semi-definite matrices with trace values $t_L$ and $t_R$ respectively: we can utilize the fact that the sum of the absolute value of matrix elements of such matrices is upper bounded by their dimensionality~\cite{horn2012matrix}, i.e. $\sum_{ij} \big|\Gamma_{ij}^{(L)}\big| \leq t_L(d_L^2 - 1),~ \sum_{ij} \big|\Gamma_{ij}^{(R)}\big| \leq t_R(d_R^2 - 1)$. Recall that $\Gamma_{ij}^{(L)}$ is a $d_L^2-1$ dimensional matrix and similarly $\Gamma_{ij}^{(R)}$ is a $d_R^2-1$ dimensional matrix.  Thus, we have from {Eq.~\eqref{b17}} that
	\begin{align} \label{b18}
		\sum_k| C_{ka}| &\leq 2(t_Ld^{{3}}_L +t_Rd^{{3}}_R - t_L{{d_L}}-t_R{{d_R}})  \notag \\
	\end{align}
	Hence, {comparing Eq.~\eqref{b18} with Eq.~\eqref{b11}}, we can {set} $\alpha = 2(t_Ld^{{3}}_L +t_Rd^{{3}}_R - t_L{{d_L}}-t_R{{d_R}})$ and arrive at Eq.~\eqref{lower_bound_maintxt} as quoted in the main text. This concludes our proof.
\end{proof}

\section{Numerical results for non-zero energy bias in half of the chain.}
\label{appn3}

We recall that in the main text, we presented numerical results for zero energy bias, i.e. the onsite energy $\omega_0$ present in Eq.~\eqref{eq:ham} was same for all sites in Figs.~\ref{fig:Nl1} and \ref{fig:Nl2}. Here we introduce a dimensionless parameter which we call the energy bias $\epsilon_0$, where $\omega^{(\ell)}_0 = 1~\text{for } \ell~\text{between 1 and } N/2$ and $\omega^{(\ell)}_0 = 1 + \epsilon_0$ for the rest. To emphasize the generality of our findings presented in Secs.~\ref{sec:nl1} and ~\ref{sec:nl2}, in this appendix, we discuss the case of non-zero energy bias.

Figs.~\ref{fig:Nl1_app}(a,b,c,d) highlight plots for the isotropic XXZ Hamiltonian for a single qubit attached to the left and right baths ($N_L = N_R = 1$) respectively with a non-zero bias of $\epsilon_0 =0.01$. Correct populations may only be possible in the equilibrium ($\beta_L = \beta_R$) scenario, while correct {populations and} coherences together are not possible in {neither} equilibrium nor non-equilibrium ($\beta_L \neq \beta_R$) setups over the parameter regimes {presented} in the plots. Figs.~\ref{fig:Nl2_app}(a,b,c,d) highlight the same type of plots but with two qubits attached to the left and right baths ($N_L = N_R = 2$). Here, correct populations may be possible but only for a very small range of values in the non-equilibrium setup as opposed to the much larger parameter range in the {zero bias} case $\epsilon_0 =0$ [see Fig.~\ref{fig:Nl2}(a)]. Correct {populations and} coherences together are still not possible for non-equilibrium setup, but is possible in the equilibrium scenario {for very low inter-site coupling strength}.

\bibliography{main.bib}
\end{document}